\PassOptionsToPackage{unicode}{hyperref}
\PassOptionsToPackage{hyphens}{url}
\PassOptionsToPackage{dvipsnames,svgnames,x11names}{xcolor}
\documentclass[
  letterpaper,
  DIV=11,
  numbers=noendperiod]{scrartcl}
\usepackage{xcolor}
\usepackage[margin=2.5cm]{geometry}
\usepackage{amsmath,amssymb}
\usepackage{iftex}
\ifPDFTeX
  \usepackage[T1]{fontenc}
  \usepackage[utf8]{inputenc}
  \DeclareUnicodeCharacter{2265}{\ensuremath{\geq}}
  \usepackage{textcomp} % provide euro and other symbols
\else % if luatex or xetex
  \usepackage{unicode-math} % this also loads fontspec
  \defaultfontfeatures{Scale=MatchLowercase}
  \defaultfontfeatures[\rmfamily]{Ligatures=TeX,Scale=1}
\fi
\usepackage{lmodern}
\ifPDFTeX\else
\fi
\IfFileExists{upquote.sty}{\usepackage{upquote}}{}
\IfFileExists{microtype.sty}{% use microtype if available
  \usepackage[]{microtype}
  \UseMicrotypeSet[protrusion]{basicmath} % disable protrusion for tt fonts
}{}
\makeatletter
\@ifundefined{KOMAClassName}{% if non-KOMA class
  \IfFileExists{parskip.sty}{%
    \usepackage{parskip}
  }{% else
    \setlength{\parindent}{0pt}
    \setlength{\parskip}{6pt plus 2pt minus 1pt}}
}{% if KOMA class
  \KOMAoptions{parskip=half}}
\makeatother
\makeatletter
\ifx\paragraph\undefined\else
  \let\oldparagraph\paragraph
  \renewcommand{\paragraph}{
    \@ifstar
      \xxxParagraphStar
      \xxxParagraphNoStar
  }
  \newcommand{\xxxParagraphStar}[1]{\oldparagraph*{#1}\mbox{}}
  \newcommand{\xxxParagraphNoStar}[1]{\oldparagraph{#1}\mbox{}}
\fi
\ifx\subparagraph\undefined\else
  \let\oldsubparagraph\subparagraph
  \renewcommand{\subparagraph}{
    \@ifstar
      \xxxSubParagraphStar
      \xxxSubParagraphNoStar
  }
  \newcommand{\xxxSubParagraphStar}[1]{\oldsubparagraph*{#1}\mbox{}}
  \newcommand{\xxxSubParagraphNoStar}[1]{\oldsubparagraph{#1}\mbox{}}
\fi
\makeatother

\usepackage{longtable,booktabs,array}
\usepackage{calc} % for calculating minipage widths
\usepackage{etoolbox}
\makeatletter
\patchcmd\longtable{\par}{\if@noskipsec\mbox{}\fi\par}{}{}
\makeatother
\IfFileExists{footnotehyper.sty}{\usepackage{footnotehyper}}{\usepackage{footnote}}
\makesavenoteenv{longtable}
\usepackage{graphicx}
\makeatletter
\newsavebox\pandoc@box
\newcommand*\pandocbounded[1]{% scales image to fit in text height/width
  \sbox\pandoc@box{#1}%
  \Gscale@div\@tempa{\textheight}{\dimexpr\ht\pandoc@box+\dp\pandoc@box\relax}%
  \Gscale@div\@tempb{\linewidth}{\wd\pandoc@box}%
  \ifdim\@tempb\p@<\@tempa\p@\let\@tempa\@tempb\fi% select the smaller of both
  \ifdim\@tempa\p@<\p@\scalebox{\@tempa}{\usebox\pandoc@box}%
  \else\usebox{\pandoc@box}%
  \fi%
}
\def\fps@figure{htbp}
\makeatother

\NewDocumentCommand\citeproctext{}{}
\NewDocumentCommand\citeproc{mm}{%
  \begingroup\def\citeproctext{#2}\cite{#1}\endgroup}
\makeatletter
 \let\@cite@ofmt\@firstofone
 \def\@biblabel#1{}
 \def\@cite#1#2{{#1\if@tempswa , #2\fi}}
\makeatother
\newlength{\cslhangindent}
\newlength{\csllabelwidth}
\newenvironment{CSLReferences}[2] % #1 hanging-indent, #2 entry-spacing
 {\begin{list}{}{%
  \setlength{\itemindent}{0pt}
  \setlength{\leftmargin}{0pt}
  \setlength{\parsep}{0pt}
  \ifodd #1
   \setlength{\leftmargin}{\cslhangindent}
   \setlength{\itemindent}{-1\cslhangindent}
  \fi
  \setlength{\itemsep}{#2\baselineskip}}}
 {\end{list}}
\usepackage{calc}

\providecommand{\tightlist}{%
  \setlength{\itemsep}{0pt}\setlength{\parskip}{0pt}}

\usepackage{booktabs}
\usepackage{longtable}
\usepackage{array}
\usepackage{multirow}
\usepackage{wrapfig}
\usepackage{float}
\usepackage{colortbl}
\usepackage{pdflscape}
\usepackage{tabu}
\usepackage{threeparttable}
\usepackage{threeparttablex}
\usepackage[normalem]{ulem}
\usepackage{makecell}
\usepackage{xcolor}
\usepackage{booktabs}
\AtBeginDocument{\author{Fabian Scheipl\thanks{All code and text were generated by large language models
  in collaboration with the author; see the statement on the use of generative AI at the end of the main text.}\\[0.6ex]
  {\small Department of Statistics, LMU Munich}\\
  {\small Munich Center for Machine Learning (MCML)}}}
\usepackage{float}
\KOMAoption{captions}{tableheading}
\makeatletter
\@ifpackageloaded{tcolorbox}{}{\usepackage[skins,breakable]{tcolorbox}}
\@ifpackageloaded{fontawesome5}{}{\usepackage{fontawesome5}}
\definecolor{quarto-callout-color}{HTML}{909090}
\definecolor{quarto-callout-note-color}{HTML}{0758E5}
\definecolor{quarto-callout-important-color}{HTML}{CC1914}
\definecolor{quarto-callout-warning-color}{HTML}{EB9113}
\definecolor{quarto-callout-tip-color}{HTML}{00A047}
\definecolor{quarto-callout-caution-color}{HTML}{FC5300}
\definecolor{quarto-callout-color-frame}{HTML}{acacac}
\definecolor{quarto-callout-note-color-frame}{HTML}{4582ec}
\definecolor{quarto-callout-important-color-frame}{HTML}{d9534f}
\definecolor{quarto-callout-warning-color-frame}{HTML}{f0ad4e}
\definecolor{quarto-callout-tip-color-frame}{HTML}{02b875}
\definecolor{quarto-callout-caution-color-frame}{HTML}{fd7e14}
\makeatother
\makeatletter
\@ifpackageloaded{caption}{}{\usepackage{caption}}
\AtBeginDocument{%
\ifdefined\contentsname
  \renewcommand*\contentsname{Table of contents}
\else
  \newcommand\contentsname{Table of contents}
\fi
\ifdefined\listfigurename
  \renewcommand*\listfigurename{List of Figures}
\else
  \newcommand\listfigurename{List of Figures}
\fi
\ifdefined\listtablename
  \renewcommand*\listtablename{List of Tables}
\else
  \newcommand\listtablename{List of Tables}
\fi
\ifdefined\figurename
  \renewcommand*\figurename{Figure}
\else
  \newcommand\figurename{Figure}
\fi
\ifdefined\tablename
  \renewcommand*\tablename{Table}
\else
  \newcommand\tablename{Table}
\fi
}
\@ifpackageloaded{float}{}{\usepackage{float}}
\floatstyle{ruled}
\@ifundefined{c@chapter}{\newfloat{codelisting}{h}{lop}}{\newfloat{codelisting}{h}{lop}[chapter]}
\floatname{codelisting}{Listing}

\makeatother
\makeatletter
\@ifpackageloaded{caption}{}{\usepackage{caption}}
\@ifpackageloaded{subcaption}{}{\usepackage{subcaption}}
\makeatother
\usepackage{bookmark}
\IfFileExists{xurl.sty}{\usepackage{xurl}}{} % add URL line breaks if available
\hypersetup{
  pdftitle={Estimation and inference in penalized function-on-function regression with within-curve residual dependence},
  pdfauthor={Fabian Scheipl},
  pdfkeywords={cluster-robust covariance, cross-validation, functional
data, generalized additive models, smoothing parameter selection},
  colorlinks=true,
  linkcolor={blue},
  filecolor={Maroon},
  citecolor={Blue},
  urlcolor={Blue},
  pdfcreator={LaTeX via pandoc}}

\title{Estimation and inference in penalized function-on-function
regression with within-curve residual dependence}
\author{Fabian Scheipl}
\date{}
\begin{document}
\maketitle
\begin{abstract}
Penalized function-on-function regression, as implemented in the R
package refund, considers all grid values as independent conditional on
the linear predictor, but some dependence typically remains. We study
the consequences for estimates and pointwise intervals in a simulation
and a plasmode study on nine real datasets. Under dependence, restricted
maximum likelihood (REML) undersmooths the coefficient surface: in our
simulations' main settings, its relative error is two to four times that
of curve-blocked neighborhood cross-validation (NCV). In our plasmode
study, both fits have large errors, possibly through weak
identification, and NCV is usually, but not always, better. Under
dependence, the independence-model-based intervals for means and
surfaces undercover severely. We propose and evaluate a curve-clustered
sandwich covariance with a leverage adjustment for the penalized hat
matrix with Satterthwaite degrees of freedom which brings coverage close
to nominal in most settings. Its intervals for function-on-function
effects are wide because REML's estimate is noisy. The same construction
at the NCV estimate is narrow but undercovers. We therefore recommend
fitting twice: NCV for the shape of the surface, REML with these
intervals for inference. For estimated means, intercept functions and
scalar covariate effects, one REML fit with these intervals usually
suffices.
\end{abstract}

\section{Introduction}\label{sec-intro}

Function-on-function regression relates a functional response \(Y_i(t)\)
to functional and scalar covariates
(\citeproc{ref-RamsaySilverman2005}{Ramsay and Silverman 2005};
\citeproc{ref-MullerYao2008}{Müller and Yao 2008};
\citeproc{ref-Morris2015}{J. S. Morris 2015}). Its terms can include,
among others, a \emph{functional intercept} \(\alpha(t)\), the linear
predictor at zero covariates; \emph{coefficient surfaces} \(\beta(s,t)\)
for functional covariates, which weight the covariate's value at
position \(s\) in its effect on the response at \(t\); and
\emph{coefficient functions} \(\gamma(t)\) for \(t\)-varying effects of
scalar covariates. In the penalized framework of Ivanescu et al.
(\citeproc{ref-IvanescuStaicuScheiplGreven2015}{2015}) and Greven and
Scheipl (\citeproc{ref-GrevenScheipl2017}{2017}), implemented in
\texttt{pffr()} of the R package refund
(\citeproc{ref-ScheiplStaicuGreven2015}{Scheipl, Staicu, and Greven
2015}; \citeproc{ref-Rpkg_refund}{Goldsmith et al. 2024}), each term is
a penalized (tensor product) spline. The model is fitted as one
generalized additive model (\citeproc{ref-Wood2017}{Wood 2017}) to the
\(N = GD\) stacked grid values of \(G\) curves with \(D\) points each,
and the fit treats these \(N\) values as independent given the additive
predictor.

Real world residual curves typically violate this assumption: they tend
to be somewhat smooth deviations from the mean, or artifacts of
misaligned features or slow drift, so values at grid points of each
curve are correlated. This dependence has two consequences. First,
smoothing-parameter selection by restricted maximum likelihood (REML),
like other data-driven selection under correlated errors
(\citeproc{ref-OpsomerWangYang2001}{Opsomer, Wang, and Yang 2001}),
undersmooths: the estimates, in particular of the coefficient surface,
are more variable than necessary. Second, the model-based standard
errors count \(N\) independent observations although the data carry much
less information. Estimated intervals are then too narrow, and finer
\(t\)-grids make them narrower without adding information.

This paper provides guidance on what users of \texttt{pffr()} (or other
functional regression software) should report, and lays out the
rationale for the defaults we have implemented. Analysts who fit such a
model require estimates that show the shape of each effect faithfully
and intervals whose error rates are close to nominal even under strong
dependence or variance heterogeneity. We evaluate the available fits and
covariances against both requirements, for Gaussian, count and binary
responses, in a simulation with 152 settings and in a plasmode study
that attaches the real residual curves of nine datasets to known
``truths''. Throughout, the model for the mean is taken to be correct
and only the dependence is misspecified; what happens when the basis
cannot represent the true effects is a separate question, which we touch
on only briefly (Section~\ref{sec-discussion}).

Curve-blocked neighborhood cross-validation (NCV)
(\citeproc{ref-Wood2024}{Wood 2024}) chooses the smoothing parameters by
leaving out one whole curve at a time. In the simulation's main
settings, REML's relative error for \(\beta(s,t)\) under dependence is
two to four times that of this NCV fit. When the true \(\beta(s,t)\) is
zero and the errors are smooth, REML's estimate shows spurious
structure, and NCV's stays much closer to zero. On the plasmode data,
the error of either fit for \(\beta(s,t)\) is large relative to the
effect, possibly through weak identification: in directions that the
functional covariate does not vary in, \(\beta(s,t)\) is free to vary
with hardly any effect on the fitted values
(\citeproc{ref-ScheiplGreven2016}{Scheipl and Greven 2016}). NCV
estimates the surface better for most ``truths'' on most datasets, but
not always.

Under dependence, model-based intervals are far too narrow for fitted
means and model terms, both in the simulation and on real residual
curves, and more so on denser grids. A curve-clustered sandwich
covariance with a leverage adjustment for the penalized hat matrix
(CL2), computed after the usual REML fit (REML + CL2) and used with
Satterthwaite degrees of freedom, brings coverage of fitted means and
\(\beta(s,t)\) close to nominal in the simulation's main settings, also
with 20 curves, and on eight of the nine plasmode datasets; it falls
short on the ninth.

Neither the REML nor the NCV fit gave both a good estimate and
calibrated intervals for the surface. REML + CL2 intervals are wide
because REML's estimate is noisy. Intervals centered at the NCV estimate
are narrow but undercover, in the simulation mainly through smoothing
bias, and the corrections we tried did not calibrate them on the
plasmode data. These results show the classical tension between
estimation and pointwise inference in nonparametric regression. An
undersmoothed estimate has small bias relative to its standard error, so
its intervals can hold their level; an estimate smoothed for accuracy
needs an explicit bias correction, whose own variability then enters the
interval (\citeproc{ref-Hall1992}{Hall 1992};
\citeproc{ref-CalonicoCattaneoFarrell2018}{Calonico, Cattaneo, and
Farrell 2018}). Here REML undersmooths through the dependence rather
than by design, and in the simulation its intervals with CL2 cover. The
tension is strongest for effects that vary along \(t\) and a second
argument. For the functional intercept, the coefficient functions of
scalar covariates and fitted means, REML + CL2 gives both a good
estimate and intervals close to nominal, except for the intercepts and
coefficient functions of binary responses and for counts under
misregistration.

Our recommendation therefore depends on the goal, which in practice may
mean fitting the model twice (Section~\ref{sec-recommend}). To learn the
shape of a \(\beta(s,t)\) surface, use curve-blocked NCV, our default,
which typically smooths more than REML. For inference about that
surface, use REML + CL2 with Satterthwaite degrees of freedom.

We make four contributions. (i) We give evidence on how much
curve-blocked NCV improves the estimate of the coefficient surface over
REML, and where it does not. (ii) We compute the CR2 leverage adjustment
of Bell and McCaffrey (\citeproc{ref-BellMcCaffrey2002}{2002}) with the
non-idempotent hat matrix of penalized fits, combine it with the
model-based smoothing-bias allowance and Satterthwaite degrees of
freedom, and implement it in the refund R package. (iii) Our evaluation
reports estimation error, width, detection of non-zero effects and false
detections of a zero effect next to coverage, in synthetic and
real-data-based settings of varying sizes and noisiness. (iv) We report
negative results on intervals centered at the NCV estimate and on other
attempted remedies in the supplement.

\emph{Related work.} Within-curve dependence can be modeled directly: by
functional mixed models with structured residual covariance
(\citeproc{ref-MorrisCarroll2006}{J. S. Morris and Carroll 2006};
\citeproc{ref-MeyerCoullVersaceCinciripiniMorris2015}{Meyer et al.
2015}), by generalized least squares with an estimated residual
covariance (\citeproc{ref-ReissHuangMennes2010}{Reiss, Huang, and Mennes
2010}) and by curve-level functional random effects
(\citeproc{ref-ScheiplGertheissGreven2016}{Scheipl, Gertheiss, and
Greven 2016}); the last two underlie two of our comparators. Signal
compression (\citeproc{ref-LuoQi2017}{Luo and Qi 2017}) is a different
estimator of the coefficient surface, and Cao, Wang, and Wang
(\citeproc{ref-CaoWangWang2020}{2020}) derive asymptotic normality for
tensor-product spline estimators of coefficient surfaces. Correlated
errors distort data-driven selection in scalar smoothing
(\citeproc{ref-OpsomerWangYang2001}{Opsomer, Wang, and Yang 2001};
\citeproc{ref-CaronParteDedeckerMichel2026}{Caron Parte, Dedecker, and
Michel 2026}); likelihood-based selection is less sensitive than
generalized cross-validation
(\citeproc{ref-KrivobokovaKauermann2007}{Krivobokova and Kauermann
2007}). Leaving out whole curves avoids modeling the dependence
(\citeproc{ref-RiceSilverman1991}{Rice and Silverman 1991}), and
neighborhood cross-validation (\citeproc{ref-Wood2024}{Wood 2024}) makes
this cheap.

For inference, pointwise and simultaneous bands for functional means and
for function-on-scalar and scalar-on-function models are well developed
(\citeproc{ref-Degras2011}{Degras 2011};
\citeproc{ref-GoldsmithGrevenCrainiceanu2013}{Goldsmith, Greven, and
Crainiceanu 2013}; \citeproc{ref-ChoiReimherr2018}{Choi and Reimherr
2018}; \citeproc{ref-LieblReimherr2023}{Liebl and Reimherr 2023}). Fast
univariate inference
(\citeproc{ref-CuiLerouxSmirnovaCrainiceanu2022}{Cui et al. 2022}) and
penalized functional generalized estimating equations
(\citeproc{ref-LoewingerLevisCuiPereira2025}{Loewinger et al. 2026})
address longitudinal functional data with many subjects. Park et al.
(\citeproc{ref-ParkStaicuXiaoCrainiceanu2018}{2018}) estimate fixed
effects under working independence and obtain their inference from a
bootstrap of subjects. Working-independence estimators for clustered
data are consistent but can be inefficient
(\citeproc{ref-LinCarroll2000}{Lin and Carroll 2000};
\citeproc{ref-WelshLinCarroll2002}{Welsh, Lin, and Carroll 2002}).
Cluster-robust covariances need only independence across clusters
(\citeproc{ref-LiangZeger1986}{Liang and Zeger 1986};
\citeproc{ref-CameronMiller2015}{Cameron and Miller 2015};
\citeproc{ref-ZeileisKoellGraham2020}{Zeileis, Köll, and Graham 2020};
\citeproc{ref-HansenLee2019}{Hansen and Lee 2019}). With few clusters,
the leverage adjustment of Bell and McCaffrey
(\citeproc{ref-BellMcCaffrey2002}{2002}) and its extensions
(\citeproc{ref-McCaffreyBell2006}{McCaffrey and Bell 2006};
\citeproc{ref-PustejovskyTipton2018}{Pustejovsky and Tipton 2018})
reduce the downward bias of the sandwich
(\citeproc{ref-KauermannCarroll2001}{Kauermann and Carroll 2001}). Here
the number of coefficients (about 100) is comparable to the number of
curves, so per-curve leverage is not negligible
(\citeproc{ref-MacKinnonNielsenWebb2023guide}{MacKinnon, Nielsen, and
Webb 2023}; \citeproc{ref-CarterSchnepelSteigerwald2017}{Carter,
Schnepel, and Steigerwald 2017}). Korolev
(\citeproc{ref-Korolev2026}{2026}) adjusts the cluster-robust covariance
for the penalty in semiparametric panel models fitted with mgcv.

\section{Model, fits and covariances}\label{sec-model}

\subsection{Model and penalized fit}\label{model-and-penalized-fit}

We work within the general functional regression framework of Greven and
Scheipl (\citeproc{ref-GrevenScheipl2017}{2017}), which relates a
feature of the conditional distribution of a functional response
\(Y_i(t)\) given covariate set \(\mathcal X_i\) to an additive
predictor, \begin{equation}\phantomsection\label{eq-gfamm}{
\boldsymbol\xi(Y_i(t) \mid \mathcal X_i) = h(\mathcal X_i) = \sum_{\ell=1}^{L} h_\ell(\mathcal X_i).
}\end{equation} The response is a function \(Y_i(t)\) over an interval
\(\mathcal T\), with scalar responses as the special case of a
single-point \(\mathcal T\), and \(\mathcal X_i\) can contain scalar and
square-integrable functional covariates as well as grouping factors. The
transformation \(\boldsymbol\xi\) determines the modeled feature: for
generalized functional additive mixed models (GFAMMs)
(\citeproc{ref-ScheiplGertheissGreven2016}{Scheipl, Gertheiss, and
Greven 2016}), the pointwise conditional mean under a link function,
\(\boldsymbol\xi = g \circ \mathrm E\). Each partial predictor
\(h_\ell(\mathcal X_i)\) is a function over \(\mathcal T\) with values
\(h_\ell(\mathcal X_i)(t)\) and depends on a subset of the covariates.
Options include functional intercepts, linear or smooth effects of
scalar and functional covariates, their interactions, and functional
random effects, in which case the model also conditions on these latent
processes.

In the following, we use a GFAMM with three partial predictors and no
functional random effects. For curves \(i = 1, \dots, G\) observed at
\(t_1, \dots, t_D\), the model is
\begin{equation}\phantomsection\label{eq-model}{
g\{\mathrm{E}[Y_i(t) \mid X_i, v_i]\} = \eta_i(t) = \alpha(t) + \int_{\mathcal S} X_i(s)\,\beta(s,t)\,ds + v_i\,\gamma(t),
}\end{equation} with link \(g\), functional intercept \(\alpha(t)\),
coefficient surface \(\beta(s,t)\) of the functional covariate
\(X_i(s)\) and coefficient function \(\gamma(t)\) of the scalar
covariate \(v_i\). A smooth effect \(f(x_i, t)\) of a scalar covariate
can replace the functional term. The responses follow an exponential
family, or an ``extended'' family such as the scaled \(t\)-distribution
(\citeproc{ref-WoodPyaSafken2016}{Wood, Pya, and Säfken 2016}), with
mean \(g^{-1}\{\eta_i(t)\}\) and variance \(\phi V(\mu)\).

Each term is expanded in a penalized spline basis, with tensor product
splines for multivariate terms. The stacked predictor is
\(\boldsymbol\eta = \mathbf X\boldsymbol\theta\) with penalty
\(\sum_r \lambda_r \boldsymbol\theta^\top \mathbf S_r \boldsymbol\theta\).
Penalized iteratively reweighted least squares gives
\(\hat{\boldsymbol\theta}\) for given smoothing parameters
\(\boldsymbol\lambda\). Let \(\mathbf W\) be the diagonal matrix of
working weights \((d\mu/d\eta)^2/\{\phi V(\mu)\}\) at convergence and
\(\mathbf S = \sum_r (\lambda_r/\phi)\mathbf S_r\). The Bayesian and the
frequentist covariance under the working model are
\begin{equation}\phantomsection\label{eq-vp}{
\mathbf V_p = (\mathbf X^\top \mathbf W \mathbf X + \mathbf S)^{-1}, \qquad
\mathbf V_e = \mathbf V_p \mathbf X^\top \mathbf W \mathbf X\, \mathbf V_p .
}\end{equation} Their difference
\(\mathbf V_p - \mathbf V_e = \mathbf V_p \mathbf S \mathbf V_p\) is a
prior-averaged allowance for smoothing bias
(\citeproc{ref-Wahba1983}{Wahba 1983}; \citeproc{ref-Nychka1988}{Nychka
1988}; \citeproc{ref-MarraWood2012}{Marra and Wood 2012}). For REML
fits, mgcv adds a term for the uncertainty of
\(\hat{\boldsymbol\lambda}\) and returns \(\mathbf V_c\)
(\citeproc{ref-WoodPyaSafken2016}{Wood, Pya, and Säfken 2016}). The
\emph{model-based} intervals below use \(\mathbf V_c\) for REML fits and
\(\mathbf V_p\) for NCV fits, because the \(\mathbf V_c\) that mgcv
returns for NCV fits in the version used here (an unpatched pre-release
build of 1.9-5) is not a valid covariance (Supplement S6). The two
model-based intervals therefore contain different uncertainty terms. All
these covariances assume that all \(N\) values \(Y_i(t_j)\) are
conditionally independent given \(\mathcal X_i\).

\subsection{Smoothing-parameter
selection}\label{smoothing-parameter-selection}

\emph{REML} maximizes the Laplace-approximate restricted likelihood of
the working model (\citeproc{ref-Wood2011}{Wood 2011}). Estimating a
term averages residual curves with weights that depend on the
covariates. Within a curve, neighboring residuals are correlated, so
pooling across grid points reduces the noise in this average much less
than assumed by the working likelihood, which counts \(N\) independent
values. We believe this excess noise, to the extent that the chosen
spline basis can represent it, is picked up as signal by REML when it
selects too little smoothing. REML's excess error grows under dependence
and on denser grids (Section~\ref{sec-res-estimation}), which fits this
reading.

\emph{Curve-blocked NCV.} Neighborhood cross-validation
(\citeproc{ref-Wood2024}{Wood 2024}) minimizes
\begin{equation}\phantomsection\label{eq-ncv}{
\mathcal C(\boldsymbol\lambda) = \sum_{k=1}^{m}\sum_{(i,j)\in \mathcal N_k} \mathrm{dev}\bigl(y_{ij}, \hat{\boldsymbol\theta}^{-\mathcal N_k}_{\boldsymbol\lambda}\bigr),
}\end{equation} the deviance of predictions for the points in
\(\mathcal N_k\) from the fit without them, with \(y_{ij} = Y_i(t_j)\).
We use one neighborhood per curve,
\(\mathcal N_k = \{(k,j): j = 1, \dots, D\}\), so \(m = G\) and each
neighborhood has as many points as its curve. The criterion then
measures how well a fit to the other curves predicts a new curve from
its covariates (\citeproc{ref-RiceSilverman1991}{Rice and Silverman
1991}). If residual curves are independent across curves, no information
about the left-out curve is contained in the noise a fit follows in the
other curves, so following its pattern worsens predictions and the
criterion selects stronger smoothing even without an explicit model of
the dependence. This holds only for the modes of variation of a term
that directly affects predictions, i.e., not for changes such as
\(\beta(s,t) + \delta(s)f(t)\) with \(\delta(s)\) in the kernel of the
covariance operator of \(X(s)\)
(\citeproc{ref-ScheiplGreven2016}{Scheipl and Greven 2016}). Where
curves are dependent on each other, the neighborhoods should be
sub-groups that are independent of each other, such as all curves of a
given subject. Spatial block neighborhoods are evaluated on the three
plasmode datasets with dependent source curves
(Section~\ref{sec-plasmode}; Supplement S2.3); subject neighborhoods
were not evaluated in this work.

No curve is actually refitted:
\(\hat{\boldsymbol\theta}^{-\mathcal N_k}\) is approximated by a single
Newton step based on the full fit, which is exact only for Gaussian
responses with identity link and given \(\boldsymbol\lambda\), and
\(\mathcal C\) is minimized over \(\boldsymbol\lambda\) with analytic
derivatives (\citeproc{ref-Wood2024}{Wood 2024}). Gaussian fits with
\(G = 100\) and \(D = 61\) using curve-blocked NCV took a median 8.6 s
against 2.0 s for REML.

\subsection{Curve-clustered covariances}\label{sec-cl2}

The curve-clustered sandwich replaces the variance part of
\(\mathbf V_p\) by one that does not assume conditional independence
within curves. It does assume conditional independence across curves as
well as a correct model for the mean. Write
\(\tilde{\mathbf X}_i = \mathbf W_i^{1/2}\mathbf X_i\) for the weighted
design rows of curve \(i\), \(\mathbf z_i\) for its Pearson residuals,
and
\(\tilde{\mathbf H} = \tilde{\mathbf X}\mathbf V_p\tilde{\mathbf X}^\top\)
for the penalized hat matrix. The CL2 covariance is
\begin{equation}\phantomsection\label{eq-cl2}{
\hat{\mathbf V}_{\mathrm{CL2}} = \frac{G}{G-1}\,\mathbf V_p \Bigl(\sum_{i=1}^G \mathbf U_i \mathbf U_i^\top\Bigr) \mathbf V_p + (\mathbf V_p - \mathbf V_e),
\qquad
\mathbf U_i = \tilde{\mathbf X}_i^\top \mathbf R_i^{-1/2}\mathbf z_i ,
}\end{equation} with
\(\mathbf R_i = [(\mathbf I - \tilde{\mathbf H})(\mathbf I - \tilde{\mathbf H})^\top]_{ii}\),
the block of curve \(i\) in the covariance of the residuals under the
working model. CR1, the usual cluster-robust covariance
(\citeproc{ref-LiangZeger1986}{Liang and Zeger 1986}), sets
\(\mathbf R_i = \mathbf I\). The adjustment \(\mathbf R_i^{-1/2}\) is
the bias-reduced linearization of Bell and McCaffrey
(\citeproc{ref-BellMcCaffrey2002}{2002}) (CR2), with the penalized hat
in place of a projection. Because \(\tilde{\mathbf H}\) is not
idempotent, \(\mathbf R_i \neq \mathbf I - \tilde{\mathbf H}_{ii}\); the
full block is computed. We keep the factor \(G/(G-1)\), floor the
eigenvalues of \(\mathbf R_i\) at \(10^{-6}\), and write CL2 to mark
these departures from CR2. The second term of Equation~\ref{eq-cl2}
keeps the model-based allowance for smoothing bias, so CL2 replaces only
the variance part of \(\mathbf V_p\). The variance part estimates the
sampling variance of \(\hat{\boldsymbol\theta}\) around its penalized
target; the added \(\mathbf V_p - \mathbf V_e\) is a prior-averaged
allowance for the distance between that target and the truth which does
not provide guaranteed coverage of every fixed truth (Appendix A.3).

The sandwich must use the penalized bread \(\mathbf V_p\) and the
penalized hat. General-purpose cluster-robust software does not provide
this combination. The clustered covariance of the sandwich package
(\citeproc{ref-Rpkg_sandwich}{Zeileis et al. 2024}) uses the penalized
bread of an mgcv fit and reproduces CR1 without the allowance, but its
leverage adjustments are restricted to unpenalized (generalized) linear
models, for whose projection hat CR2 is also defined in clubSandwich
(\citeproc{ref-Rpkg_clubSandwich}{Pustejovsky 2024}). Limited asymptotic
consistency and validity results for fixed smoothing parameters are
provided in Appendix A.5, but in the realistic regimes we study here
(about as many spline coefficients as curves, non-negligible penalties
and estimated smoothing parameters), the required assumptions do not
apply. CL2 is a finite-sample procedure whose behavior is established
empirically below.

\emph{Satterthwaite degrees of freedom.} For any contrast
\(\mathbf a^\top\boldsymbol\theta\), the variance part of
Equation~\ref{eq-cl2} is a sum of squares
\(\sum_i (\mathbf q_i^\top\mathbf z_i)^2\) with
\(\mathbf q_i = \mathbf R_i^{-1/2}\tilde{\mathbf X}_i\mathbf V_p\mathbf a\),
up to the factor \(G/(G-1)\). Under a Gaussian working model with fixed
\(\boldsymbol\lambda\) and no smoothing bias, the residuals are
\((\mathbf I - \tilde{\mathbf H})\boldsymbol\varepsilon\) with
independent standard normal \(\boldsymbol\varepsilon\), and matching the
first two moments of the quadratic form to a scaled \(\chi^2_\nu\) gives
\(\nu = (\operatorname{tr}\boldsymbol\Gamma)^2/\operatorname{tr}(\boldsymbol\Gamma^2)\)
with
\(\Gamma_{ih} = \mathbf q_i^\top[(\mathbf I - \tilde{\mathbf H})(\mathbf I - \tilde{\mathbf H})^\top]_{ih}\mathbf q_h\)
(see Appendix A.7 and \citeproc{ref-BellMcCaffrey2002}{Bell and
McCaffrey 2002}; \citeproc{ref-PustejovskyTipton2018}{Pustejovsky and
Tipton 2018}). The interval uses the \(t_\nu\) quantile at each grid
point. We use the same \(\nu\) for all families (Appendix A.7 discusses
this choice). The degrees of freedom ignore non-normality of the
responses, the bias allowance, the dependence and the selection of
\(\boldsymbol\lambda\); they are a reference distribution, not an exact
pivot.

The same constructions apply to both fits: \emph{REML + CL2} is
Equation~\ref{eq-cl2} at the REML estimate and \emph{NCV + CL2} at the
NCV estimate.

\subsection{Estimands and performance measures}\label{sec-measures}

Four estimands are evaluated on fixed grids: the conditional mean
\(\mathrm E(Y \mid X)\) for a fixed set of 50 subjects' covariates, on
the response scale; \(\beta(s,t)\) (or \(f(x,t)\)); \(\alpha(t)\); and
\(\gamma(t)\), on the link scale. Pointwise intervals are estimate
\(\pm\) critical value \(\times\) standard error, transformed by
\(g^{-1}\) for the mean. Every method is judged on accuracy of the
estimate, calibration and informativeness
(\citeproc{ref-MorrisWhiteCrowther2019}{T. P. Morris, White, and
Crowther 2019}):

\begin{itemize}
\tightlist
\item
  \emph{Coverage}: the share of replicates whose interval contains the
  truth at a grid point, averaged over the \(J\) evaluation points of
  the estimand. Its Monte Carlo standard error is the standard deviation
  over replicates of the grid-averaged coverage divided by the square
  root of the number of replicates. The \emph{lower tail} is the 5\%
  quantile of pointwise coverage over the grid. Each pointwise coverage
  is itself estimated from 200 replicates, so the lower tail lies below
  the nominal level even for a perfectly even interval: with coverage
  0.95 at every point, it would be about 0.925.
\item
  \emph{Relative error}:
  \(\{\sum_j(\hat\vartheta_j - \vartheta_j)^2/\sum_j\tilde\vartheta_j^2\}^{1/2}\)
  over grid points \(j\), for estimate \(\hat\vartheta\) and truth
  \(\vartheta\), with \(\tilde\vartheta = \vartheta\) for \(\beta\) and
  \(\gamma\) and \(\tilde\vartheta\) the truth centered at its grid mean
  for the mean and \(\alpha(t)\), whose level is not the effect. A value
  of 1 means the error is as large as the effect. We report its median
  over replicates.
\item
  \emph{Relative half-width}: the root-mean-square half-width of the
  intervals over the grid, divided by
  \((\sum_j\tilde\vartheta_j^2/J)^{1/2}\), the root-mean-square size of
  the (centered) truth.
\item
  \emph{Detection rate}: among grid points where the absolute true value
  exceeds 25\% of its maximum,
  \(|\vartheta_j| > 0.25\max_k|\vartheta_k|\), the share whose interval
  excludes zero on the side of the truth.
\item
  \emph{Interval score} (\citeproc{ref-GneitingRaftery2007}{Gneiting and
  Raftery 2007}): for an interval \([l, u]\) and true value
  \(\vartheta\),
  \((u - l) + (2/0.05)\{(l - \vartheta)_+ + (\vartheta - u)_+\}\),
  averaged over the grid and the replicates; lower is better. It is a
  proper scoring rule for predictive quantile intervals; against fixed
  truths over repeated datasets it rewards an accurate center and a
  narrow width together and is not a calibration criterion. We report it
  relative to REML + CL2 and interpret it in context of the other
  measures.
\end{itemize}

The supplement adds wrong-sign rates and studentized errors. Comparisons
of estimators use the ratio of the two fits' median relative errors;
where ratios of mean squared errors are given, they are labeled as such.
In Section~\ref{sec-results}, both are REML to NCV.
Table~\ref{tbl-constructions} lists the principal interval constructions
and where each was evaluated.

\begin{table}

\caption{\label{tbl-constructions}Principal interval constructions.
Covariance, bias-allowance and critical-value variants are specified in
Appendix A.6; comparator and NCV-centered constructions in Supplement S3
and S4. All CL2 variants cluster by curve, use the penalized bread
\(\mathbf V_p\) and include the allowance \(\mathbf V_p - \mathbf V_e\)
unless called frequentist. Full block: \(\mathbf R_i\) of Equation 5;
shortcut: \(\mathbf R_i \approx \mathbf I - \tilde{\mathbf H}_{ii}\).}

\centering{

\centering\begingroup\fontsize{8}{10}\selectfont

\begin{tabular}{lllll}
\toprule
interval & fit & covariance & critical value & evaluated in\\
\midrule
model-based & REML & mgcv's $\mathbf V_c$ & z & all settings, plasmode\\
model-based & NCV & $\mathbf V_p$ & z & all settings (supplement)\\
REML + CL2 & REML & CL2, full block, + $(\mathbf V_p - \mathbf V_e)$ & z & all settings, plasmode\\
REML + CL2, Satterthwaite & REML & as REML + CL2 & $t_\nu$ (Satterthwaite) & main factorial (default signal), extensions, plasmode\\
NCV + CL2 & NCV & CL2, full block, + $(\mathbf V_p - \mathbf V_e)$ & z & all settings, plasmode\\
CR1, CL2 shortcut, frequentist CL2, $t_{G-1}$ & REML & variants of CL2 & $z$ or $t_{G-1}$ & main factorial, default signal\\
\bottomrule
\end{tabular}
\endgroup{}

}

\end{table}%

\section{Simulation design}\label{sec-design}

The simulation follows the model in Equation~\ref{eq-model} with a
functional covariate whose spectrum matches the corpus-callosum profiles
of the DTI data in refund (152 settings, 200 replicates each, 30,400
fits per selector; Appendix B.1 lists all blocks of settings). The
truths are fixed smooth functions, projected onto the used spline bases
so that they are representable without error; ``rough'' variants add
higher-frequency components. Responses are Gaussian (signal-to-noise
\(R^2 = 0.2, 0.5, 0.8\)), Poisson and Bernoulli at each grid point (two
signal-to-noise levels on the link scale), with scaled \(t\), Beta and
Negative-Binomial responses as extensions.

Within-curve errors are zero-mean processes with correlation kernel
\(R\), used directly for Gaussian responses and as the latent layer of a
Gaussian copula otherwise, so that every marginal remains in the fitted
family. So for count and binary data, the dependence is a copula
construction, and the support for CL2 with these families is empirical
within that construction. Three error processes form the main factorial:

\begin{itemize}
\tightlist
\item
  \emph{independent}: \(R(t,t') = \mathbb 1\{t = t'\}\);
\item
  \emph{rough stationary}:
  \(R(t,t') = 0.8\exp(-|t-t'|/0.089) + 0.2\,\mathbb 1\{t=t'\}\);
\item
  \emph{smooth non-stationary}: the residual correlation of the DTI
  regression in refund (Figure~\ref{fig-kernels} in Appendix B.1).
\end{itemize}

The main factorial crosses these processes with the family,
\(G \in \{40, 100\}\) and the signal level at \(D = 61\) grid points.
Each extension changes one aspect of this design (Appendix B.1 lists the
settings of every block): the main factorial with \(G = 20\), a four
times denser grid, a rough truth, a coefficient surface that is zero
everywhere (\(\beta \equiv 0\); Gaussian and binary responses,
independent and smooth errors), a smooth effect \(f(x,t)\) in place of
the functional term \(\int X(s)\beta(s,t) ds\), sign-changing and
heteroskedastic errors, misregistration (each curve observed on its own
clock / deformed elastically), a low-rank functional covariate and
larger bases.

The fitted bases are cubic B-splines with first-order difference
penalties (mgcv's \texttt{ps} basis with \texttt{m\ =\ c(2,\ 1)}), so
the penalty null space contains the constants along each direction:
\(k = 12\) along \(t\) for \(\alpha(t)\) and \(\gamma(t)\) and an
\(8 \times 10\) tensor product for \(\beta(s,t)\), about 100
coefficients in total; the larger bases halve and quarter the knot
spacing. Results with second-order penalties were not computed.

Each replicate is fitted by REML and by curve-blocked NCV; none of the
fits failed. For NCV fits of scaled-\(t\) responses, mgcv often reported
that the search over the smoothing parameters ended without meeting its
convergence test (88\% under smooth errors, 68\% under independent
errors, but results were mostly stable under refits from different
starting values indicating flat optima), and rarely in other families;
REML fits were not affected (settings with at least 40 curves and a
non-zero surface; Appendix B.1). Random draws depend only on the
replicate, so contrasts between methods and settings are paired. The
Monte Carlo standard error of a grid-averaged coverage is 0.21--0.33 pp
(percentage points) for REML + CL2 in the dependent main-factorial
settings with 100 curves.

\section{Simulation results}\label{sec-results}

\begin{figure}

\centering{

\pandocbounded{\includegraphics[keepaspectratio]{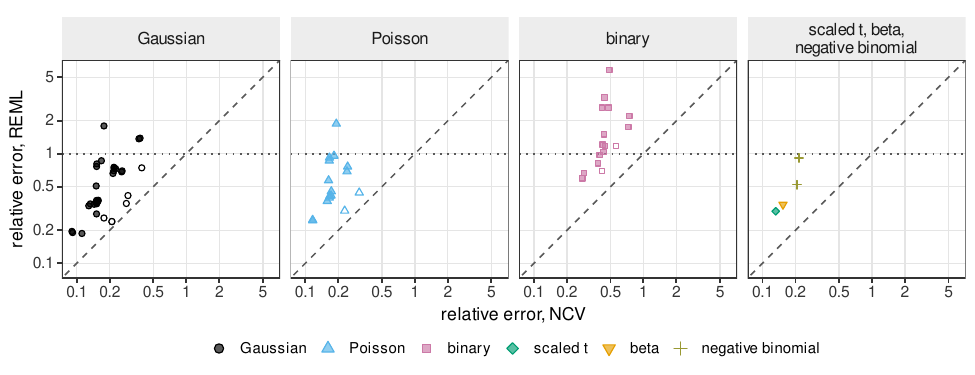}}

}

\caption{\label{fig-estimation}Relative error of the REML and the NCV
estimate of \(\beta(s,t)\) in all simulation settings with dependent
errors, 40 or 100 curves and a non-zero surface (one symbol per setting;
panels by family on common axes; median over replicates; log scales;
open symbols: rough truth; dashed: equal error; dotted: error as large
as the effect).}

\end{figure}%

\subsection{Estimation: REML versus NCV}\label{sec-res-estimation}

Under independent errors, both selectors estimate \(\beta(s,t)\) well:
the median relative error over the settings is 0.16 for REML and 0.12
for NCV. Under dependence, REML's error rises to 0.67 (maximum 5.8) and
NCV's to 0.21 (Figure~\ref{fig-estimation}). In the main factorial,
REML's median relative error is 2.1--3.7 times NCV's under dependence,
at every signal level (Table~\ref{tbl-ratio}). The ratio is larger on a
denser grid and with larger bases, and smaller with a rough truth and
for a smooth effect \(f(x,t)\) in place of the functional term. NCV
estimates \(\beta(s,t)\) better also under independent errors when the
truth is smooth (ratio of mean squared errors 1.1--1.8 for Gaussian and
Poisson responses; 0.87--1.02 for binary responses), but not when it is
rough (0.85--0.98 with 100 curves). Part of the gap therefore reflects
the smooth truths rather than the dependence: REML selects the smoothing
parameters under the prior implied by the penalty, and undersmoothing a
truth that is smoother than this prior is a known property of
likelihood-based selection (\citeproc{ref-Wahba1985}{Wahba 1985}).
Leaving out single grid points instead of whole curves does not give
NCV's gain (Supplement S1.5).

\begin{table}

\caption{\label{tbl-ratio}Ratio of the median relative errors of the
REML and the NCV estimate of \(\beta(s,t)\) (REML to NCV; smooth effect:
\(f(x,t)\)): range over the settings of each block. The extensions have
dependent errors unless stated; misregistered curves have independent
errors.}

\centering{

\centering\begingroup\fontsize{8}{10}\selectfont

\begin{tabular}{ll}
\toprule
settings & REML / NCV\\
\midrule
main factorial, dependent errors & 2.1–3.7\\
main factorial, independent errors & 1.1–1.4\\
20 curves, dependent errors & 2.5–4.7\\
20 curves, independent errors & 1.0–1.3\\
four times denser grid & 5.0–7.4\\
larger bases & 5.2–12.0\\
rough truth & 1.3–2.1\\
misregistration & 1.1–1.9\\
smooth effect in place of the functional term & 1.3–1.7\\
\bottomrule
\end{tabular}
\endgroup{}

}

\end{table}%

REML's surface uses 1.9--2.3 times NCV's effective degrees of freedom
(EDF) under dependence (1.1--1.3 under independence). Its excess error
is variance, not bias: squared bias is 0.01 of REML's mean squared error
for \(\beta\) and 0.12--0.20 of NCV's (100 curves, smooth truth). Larger
bases make it worse: with the largest basis, REML's surface EDF under
smooth errors grows from 27--42 to 113--176 across families, NCV's from
12--21 to 18--41 (Supplement S5).

Estimates for the mean, the intercept and the scalar effect depend much
less strongly on the selector. Under independent errors, the median
relative error over the settings is 0.11 (REML) and 0.10 (NCV) for the
mean, 0.16 and 0.16 for \(\alpha(t)\), and 0.11 and 0.12 for
\(\gamma(t)\). Under dependence, it is 0.28 and 0.23, 0.31 and 0.28, and
0.22 and 0.22 (Figure~\ref{fig-tradeoff-summary}). In the dependent
main-factorial settings, the ratio of REML's to NCV's median relative
error is 1.1--1.3 for the mean, 0.94--1.30 for \(\alpha(t)\) and
0.98--1.13 for \(\gamma(t)\), against 2.1--3.7 for \(\beta\). NCV thus
gains modestly for the mean (ratio of mean squared errors 1.2--1.7 with
100 curves) and little for the other two terms; for \(\alpha(t)\) the
ratio stays at or below 1.09 outside binary responses. REML also has the
heavier tail: 4.3\% of replicates have a relative error above 1 for the
mean, against 0.7\% for NCV, with smaller differences for \(\alpha(t)\)
(10.8\% and 8.5\%) and \(\gamma(t)\) (3.8\% and 3.2\%).

\emph{Stability of the smoothing parameters.} GCV can have multiple
minima and more variable smoothing parameters than REML
(\citeproc{ref-ReissOgden2009}{Reiss and Ogden 2009}), and
prediction-error criteria occasionally undersmooth severely
(\citeproc{ref-Wood2011}{Wood 2011}). Curve-blocked NCV, which these
results do not cover, shows a similar tail. With 40 or 100 curves
(Appendix B.1), the first of the surface's two smoothing parameters fell
more than a factor of \(e^3 \approx 20\) below its setting's median in
2.2\% of NCV fits and in no REML fit under independent errors, and REML
estimated the surface better on 99\% of those replicates. Such fits made
up 6.0\% of NCV fits under dependence, apart from misregistration, but
there NCV still estimated the surface better on 97\% of those
replicates. They have 2.6 times their setting's median relative error
and account for 31\% of NCV's squared error, but they do not explain its
undercoverage: leaving them out moves NCV + CL2 only from 0.900 to
0.904. Convergence flags do not detect them: outside scaled-\(t\)
responses, only 1 of 542 was flagged.

Neither fit is uniformly the more stable under dependence: the median
standard deviations of the two log smoothing parameters over replicates
are 1.34 and 0.83 for NCV and 2.09 and 0.37 for REML (under
independence, 0.91 and 0.75 against 0.14 and 0.10). Restarts of flagged
scaled-\(t\) fits can sometimes result in materially different fits
(Appendix B.1); multiple starts were not evaluated as a fitting
procedure.

\emph{Alternative models.} Explicit models for residual dependence also
reduce REML's surface estimation error. Two such methods were compared
with NCV on the same data, in 4 Gaussian main-factorial settings with
dependent errors as well as 12 plasmode combinations (Supplement S3.2):
a functional random effect for each curve based on the eigenfunctions of
the REML-estimated residual curves (\texttt{pcre()} in refund), and
generalized least squares (GLS) with a residual covariance estimated
from the REML residual curves by functional principal components (FPCA)
plus white noise (GLS-FPCA). Their mean squared error for
\(\beta(s,t)\), relative to NCV's (geometric mean over settings), is
1.75 (pcre) and 1.79 (GLS) in the simulation, where REML's is 7.4, and
0.72 and 0.68 on the plasmode data (midpoint truth, defined in
Section~\ref{sec-plasmode}, with REML residuals; the GLS covariance is
estimated from the residual curves of a REML pilot fit to each simulated
dataset, and because these errors are sign-flipped REML residuals of the
original fit, those pilot residuals closely resemble the curves that
generate them). Both are thus much better than REML, worse than NCV in
the simulation and comparable to NCV in that plasmode comparison.
Including the REML pilot fit, a pcre fit took 2.3--20 times as long as
an NCV fit in the simulation and 0.9--20 times as long on the plasmode
data; a GLS fit took 0.6--0.7 and 0.4--0.9 times as long (medians per
setting). The GLS algorithm -- only applicable for Gaussian data -- is
given in Supplement S3.2.

\begin{figure}

\begin{minipage}{\linewidth}

\centering{

\pandocbounded{\includegraphics[keepaspectratio]{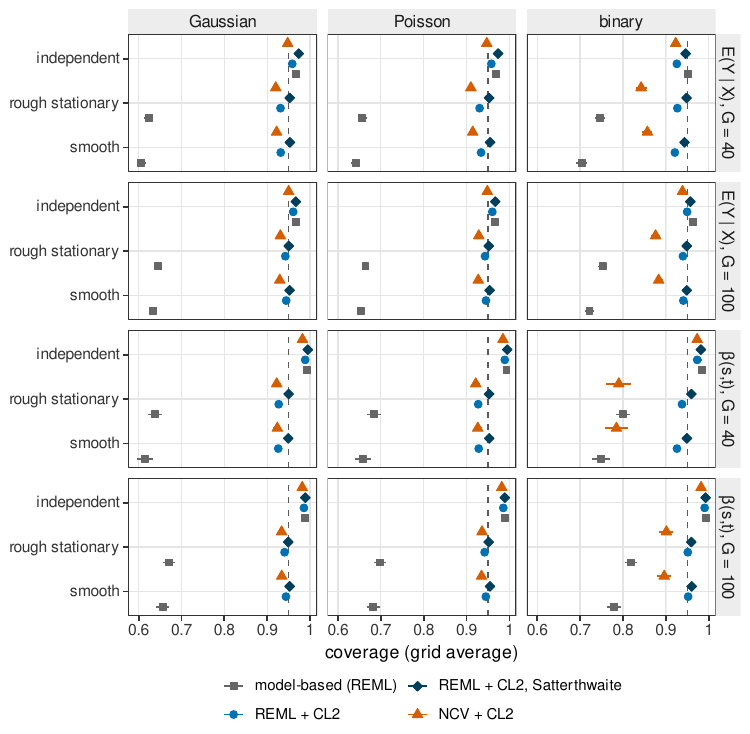}}

}

\subcaption{\label{fig-coverage}Coverage of the mean and of
\(\beta(s,t)\)}

\end{minipage}%
\newline
\begin{minipage}{\linewidth}

\centering{

\pandocbounded{\includegraphics[keepaspectratio]{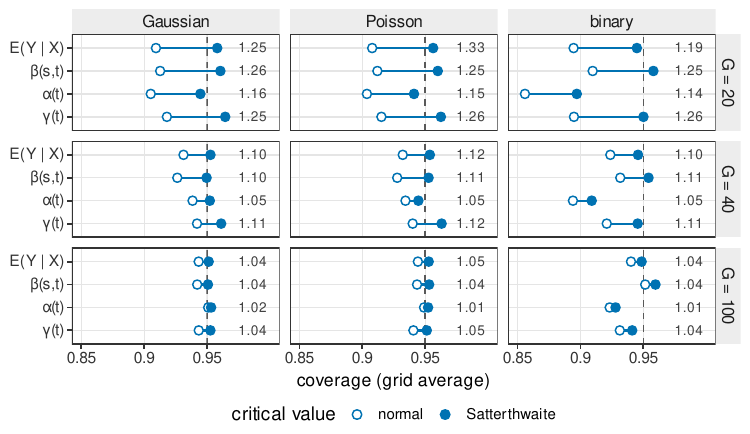}}

}

\subcaption{\label{fig-satt}Normal and Satterthwaite critical values for
REML + CL2}

\end{minipage}%

\caption{\label{fig-intervals}Grid-averaged coverage of nominal 95\%
pointwise intervals, main factorial at the default signal level (dashed:
0.95). (a) By family (columns), error process, estimand and number of
curves \(G\) (rows). REML + CL2 and NCV + CL2 use the normal critical
value; REML + CL2, Satterthwaite, is the recommended interval. Bars:
\(\pm 2\) Monte Carlo SE. NCV + CL2 is discussed in
Section~\ref{sec-res-tradeoff}. Coverage for \(\alpha(t)\) and
\(\gamma(t)\): Figure~\ref{fig-s1-alpha-gamma} in Appendix A.6; summary
over all settings: Figure~\ref{fig-tradeoff-summary}. (b) Dependent
errors (mean over the two processes). Numbers at the right: width of the
Satterthwaite interval relative to the normal one.}

\end{figure}%

\subsection{Intervals for the REML fit}\label{sec-res-intervals}

Model-based intervals undercover in every dependent setting and every
family (Figure~\ref{fig-coverage}): 0.63--0.82 for the mean and
\(\beta(s,t)\) with 100 curves. On a four times denser grid
(\(D = 241\)), their coverage of \(\beta\) drops to 0.35--0.45. REML +
CL2 covers the mean and \(\beta\) at 0.94--0.95 with 100 curves and
stays at 0.94 for \(\beta\) at \(D = 241\). The same holds for
\(\alpha(t)\) and \(\gamma(t)\) of Gaussian and count responses, for
scaled \(t\), beta and negative-binomial responses, for the
smooth-effect model, for sign-changing and heteroskedastic errors, and
for misregistered Gaussian curves (Supplement S1.1--S1.3; misregistered
Gaussian curves: mean and \(\beta\) 0.93--0.95). In the dependent
main-factorial settings, REML + CL2's coverage is also much more even
over the grid: the lower tail for the mean and \(\beta\) is 0.91--0.93
with 100 curves and 0.89--0.91 with 40, against 0.54--0.76 for the
model-based intervals with 100 curves.

Under independent errors with 100 curves, CL2 gives up 0.3--1.1 pp of
coverage against the model-based intervals for the mean and \(\beta\)
(averaged over signal levels within each family). For binary responses
with 40 curves at the default signal, it gives up 2.6 pp for the mean,
where the model-based intervals do not overcover (REML + CL2 0.93,
model-based 0.95), and 4.4 pp and 4.1 pp for \(\alpha(t)\) and
\(\gamma(t)\) (Appendix A.6). For \(\beta(s,t)\), the intervals
overcover under independent errors (Figure~\ref{fig-coverage}), because
the allowance exceeds the smoothing bias of a smooth truth (Appendix
A.3): REML + CL2 with the Satterthwaite critical value covers 0.98--0.99
(Appendix A.7).

\emph{Fewer curves.} With 40 curves, REML + CL2 covers the mean and
\(\beta\) at 0.92--0.94. With 20 curves, it covers the mean, \(\beta\)
and \(\gamma\) at 0.88--0.93 and \(\alpha(t)\) at 0.83--0.91.
Satterthwaite degrees of freedom recover most of this shortfall
(Figure~\ref{fig-satt}): with 20 curves and the default signal, coverage
of the mean, \(\beta\) and \(\gamma\) rises from 0.89--0.92 to
0.94--0.96, at up to 1.33 times the width; with 100 curves, the change
is small. The binary intercept stays short (0.90 with 20 curves, 0.91
with 40). The degrees of freedom are about a third of \(G - 1\) for the
mean, \(\beta\) and \(\gamma\), and larger for \(\alpha(t)\); Appendix
A.7 explains why. In the extensions, they never lower coverage. There,
with Satterthwaite degrees of freedom, coverage stays below 0.93 in 13
of 291 combinations of setting and estimand, all under misregistration
or for binary intercepts and scalar effects (Supplement S1.7).

\emph{CL2 and CR1.} Under dependence and with the normal critical value,
CL2 covers the mean and \(\beta\) 3.1--4.2 pp more than CR1 with the
same allowance with 40 curves, at 1.1--1.2 times its width, and 1.0--1.4
pp more with 100, at 1.04--1.06 times (Appendix A.6). For fixed
smoothing parameters and a fixed basis, the two covariances agree to
first order as the number of curves grows (Appendix A.5). CR1 was not
run with 20 or with more than 100 curves. The leverage adjustment costs
a median of 1.0 s per fit with 100 curves.

\emph{Where CL2 falls short.} For binary responses, REML + CL2 covers
the functional intercept and the scalar effect at 0.88--0.94 under
dependence and 0.86--0.95 under independent errors. Under independent
errors with 40 curves, the intercept's studentized error is centered but
too spread out, which is compatible with a standard error that is too
small, although local biases of opposite sign are not excluded. Under
dependence, it is also shifted (Supplement S1.4). Fixing the smoothing
parameters at their typical value raises coverage of the binary
\(\gamma(t)\) under independent errors from 0.89 to 0.93 and of
\(\alpha(t)\) under smooth errors from 0.88 to 0.92 (40 curves), so part
of the shortfall is likely due to uncounted smoothing-parameter
variability; \(\alpha(t)\) under independent errors stays at 0.87
(Appendix A.6, Supplement S1.4). With 20 curves and independent errors,
every interval undercovers binary \(\alpha(t)\) and \(\gamma(t)\), the
model-based ones included (0.80--0.93), so this shortfall is not
specific to the sandwich.

For misregistered counts, the working model for the mean is wrong: REML
+ CL2 covers the mean of misregistered Poisson responses at 0.90--0.91
(model-based 0.56--0.63).

\subsection{The trade-off for the coefficient
surface}\label{sec-res-tradeoff}

In the simulation, REML's estimate of \(\beta(s,t)\) is noisy but has
little bias, and CL2 measures its spread, so its intervals cover. NCV's
estimate has smaller error but more bias, beyond what the allowance
\(\mathbf V_p - \mathbf V_e\) covers, so its narrower intervals cover
less. A more accurate estimate therefore need not give better coverage.

\begin{figure}

\begin{minipage}{\linewidth}

\centering{

\pandocbounded{\includegraphics[keepaspectratio]{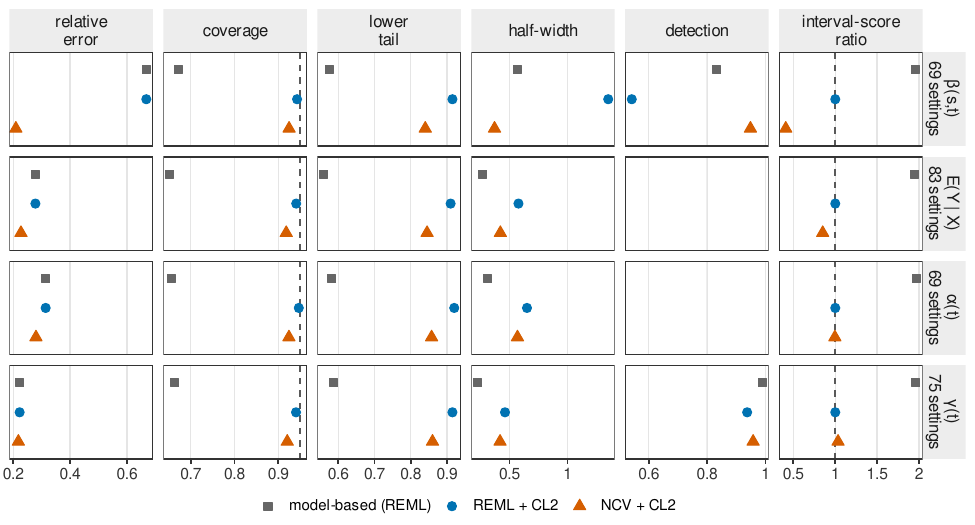}}

}

\subcaption{\label{fig-tradeoff-summary}Summary per estimand and
interval}

\end{minipage}%
\newline
\begin{minipage}{\linewidth}

\centering{

\pandocbounded{\includegraphics[keepaspectratio]{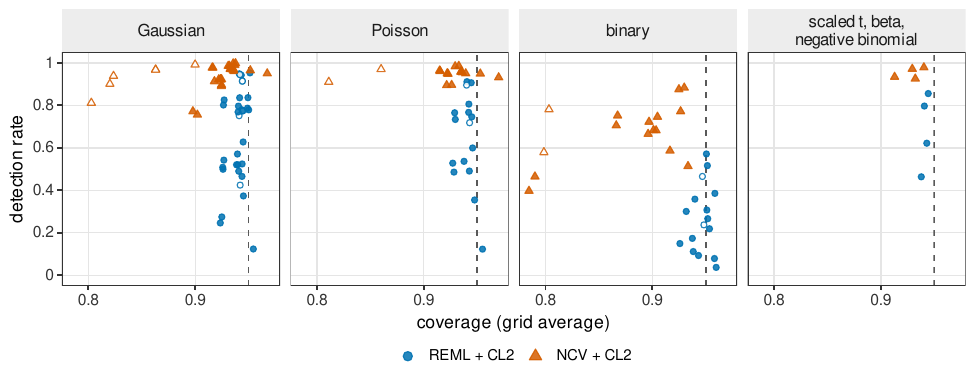}}

}

\subcaption{\label{fig-tradeoff}Coverage and detection rate for
\(\beta(s,t)\) per setting}

\end{minipage}%

\caption{\label{fig-tradeoff-all}All simulation settings with dependent
errors, 40 or 100 curves and a non-zero surface (dashed: 0.95 for
coverage, 1 for the ratio). (a) Median over settings of the relative
error of the estimate, the grid-averaged coverage, the lower tail (5\%
quantile of pointwise coverage), the relative half-width and the
detection rate, and the interval score relative to REML + CL2 (geometric
mean over settings). The number of settings differs by estimand:
\(\beta(s,t)\) and \(\alpha(t)\) are not scored in the smooth-effect
model, and only the mean is scored for misregistered count and binary
responses. Detection is defined for \(\beta(s,t)\) and \(\gamma(t)\)
only. Values: Supplement S7. (b) One symbol per setting and interval;
panels by family on common axes; open symbols: rough truth.}

\end{figure}%

Under dependence, the median half-width of REML + CL2 for \(\beta\) is
1.4 times the size of the effect, and it detects only 0.54 of the grid
points where the true value exceeds 25\% of its maximum
(Figure~\ref{fig-tradeoff-summary}, Figure~\ref{fig-tradeoff}). Its
intervals are 3.7 times as wide as those of NCV + CL2, which detects
0.95 but covers less (median 0.92, lower tail 0.84). These widths and
detection rates are for the normal critical value. In the main factorial
at the default signal, the Satterthwaite critical value widens REML +
CL2 by a further factor of 1.04--1.11 (Figure~\ref{fig-satt}). Its
detection rate for \(\beta(s,t)\) under dependence then falls by at most
6 pp, to 0.11--0.49 with 40 curves and 0.25--0.78 with 100. The interval
score of NCV + CL2 is 0.41 times REML + CL2's: the score rewards its
accurate, narrow intervals more than it punishes the additional misses.
We nevertheless do not recommend NCV + CL2 for inference, because its
coverage is too low on the real residual curves
(Section~\ref{sec-plasmode}).

In the simulation, the shortfall of NCV + CL2 is mainly smoothing bias:
subtracting the bias of the NCV estimate, estimated from the replicates,
raises its coverage from 0.90--0.94 to 0.93--0.96 in the main factorial
with 100 curves. NCV + CL2 covers less with 20 curves and dependent
errors (0.71--0.92 for the four estimands), with a rough truth
(0.80--0.90; REML + CL2: 0.94--0.95) and for binary responses with 40
curves (0.78--0.90).

The trade-off grows with the basis: with the largest basis and smooth
errors, REML + CL2's half-width for \(\beta\) reaches 4--12 times the
effect and its detection rate falls to 0.04--0.12, while its coverage
stays at 0.95--0.96. Wrong-sign exclusions are rare for REML + CL2 (at
most 2.3\% of the grid in any dependent setting).

For the mean, \(\alpha(t)\) and \(\gamma(t)\), the trade-off is mild:
REML + CL2's intervals are wider than NCV + CL2's (mean: 1.4 times), but
they cover, the estimates of \(\alpha(t)\) and \(\gamma(t)\) are about
as accurate, and \(\gamma(t)\) is detected at 0.94.

The model-based intervals of pcre and GLS with an FPCA residual
covariance show the trade-off from the other side. In the 4 Gaussian
main-factorial settings with dependent errors, they cover \(\beta\) at
0.93--0.97, at 0.5--0.7 times REML + CL2's width and about half its
interval score, but pcre's intervals cover the mean at only 0.88--0.91
and \(\alpha(t)\) at 0.51--0.59. On real residual curves, both cover
\(\beta\) at only 0.59--0.86 (Section~\ref{sec-plasmode}).

CL2 also applies to these fits: GLS whitens the data within curves, so
curves stay independent clusters, and pcre + CL2 is CL2 on the pcre fit
(Supplement S3.2). Under dependent synthetic errors, GLS-FPCA + CL2
covers \(\beta\) at 0.96, at 0.6 times REML + CL2's width, and pcre +
CL2 repairs pcre's intervals for \(\alpha(t)\) (0.56 to 0.95). On the
real residual curves, GLS-FPCA + CL2 covers \(\beta\) at 0.82
(0.65--0.90 by dataset) and pcre + CL2 at 0.83, against 0.93 for REML +
CL2. GLS with the raw residual covariance fails with or without CL2
(0.65 under dependent synthetic errors). Both pcre and GLS-FPCA thus
estimate better than REML (Section~\ref{sec-res-estimation}), and the
sandwich repairs much, but not all, of their undercoverage.

\subsection{A zero coefficient surface}\label{sec-res-null}

\begin{figure}

\centering{

\pandocbounded{\includegraphics[keepaspectratio]{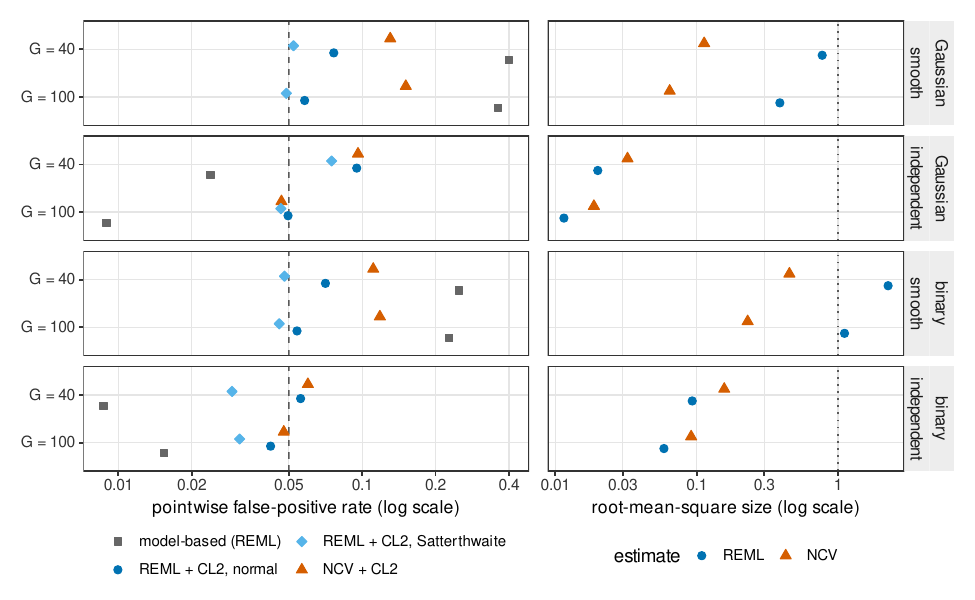}}

}

\caption{\label{fig-null}Zero surface (\(\beta \equiv 0\)), default
signal, by family, errors and number of curves \(G\) (log scales). Left:
pointwise false-positive rate of the model-based REML interval, REML +
CL2 with the normal and the Satterthwaite critical value and NCV + CL2
(dashed: nominal 0.05). Right: root-mean-square size of the REML and NCV
estimates relative to the non-zero surface of the corresponding setting
(dotted: 1). Values: Supplement S1.8 and S7.}

\end{figure}

With \(\beta \equiv 0\), the share of grid points at which an interval
excludes zero is the pointwise false-positive rate, nominally 0.05
(Figure~\ref{fig-null}). Under smooth errors, it is 0.05--0.08 for REML
+ CL2 with the normal critical value and 0.05 with Satterthwaite degrees
of freedom, against 0.11--0.15 for NCV + CL2 and 0.23--0.57 for the
model-based intervals of either fit. Under independent errors, REML +
CL2 has 0.04--0.06, except 0.095 for Gaussian responses with 40 curves
(0.075 with the Satterthwaite critical value; NCV + CL2: 0.096). There,
REML shrinks the surface to near the null space of its penalty in most
replicates, and in some of these the interval excludes zero over much of
the grid, a known weakness of intervals for terms shrunk to their null
space (\citeproc{ref-MarraWood2012}{Marra and Wood 2012}).

Under smooth errors, REML's estimate of the zero surface has 0.4--2.3
times the root-mean-square size of the non-zero surface in the
corresponding setting: spurious structure. NCV's estimate is 5--7 times
smaller. Under independent errors, both are close to zero.

Pointwise intervals do not answer whether \(\beta(s,t)\) is zero
everywhere: under smooth errors, 0.72--0.83 of the replicates have at
least one grid point where the Satterthwaite interval excludes zero, and
0.33--0.41 have more than 5\% of the grid excluded. That question needs
simultaneous bands or global tests, which this study does not evaluate.

\subsection{Other intervals and fits}\label{sec-res-other}

We also evaluated the following constructions.

\begin{itemize}
\tightlist
\item
  \emph{Intervals centered at the NCV estimate.} We tried widening by
  the NCV--REML difference, the NCV estimate with REML + CL2's
  half-width, a one-step bias correction of the NCV estimate and a
  cluster-robust term for the variability of the selected smoothing
  parameters, alone and combined. The combination is a frozen-map
  approximation: it propagates the smoothing-parameter term through the
  bias-correction map but ignores how that map changes with the
  smoothing parameters. In Gaussian settings, it covers \(\beta(s,t)\)
  at 0.97 (smooth truths) and 0.94 (rough truths), but none of these
  constructions held its level across the plasmode datasets
  (Section~\ref{sec-plasmode-results}; Supplement S4.1).
\item
  \emph{mgcv's neighborhood covariance for NCV fits.} For curve blocks,
  the unpatched mgcv build used here returns the Bayesian
  \(\mathbf V_p\) or, with \texttt{jackknife\ =\ TRUE}, essentially
  \(\mathbf V_p - \mathbf V_e\); it computes this covariance correctly
  only for per-point neighborhoods. Computed correctly from the block
  fit, it is conservative in the simulation for \(\beta\) (0.96--0.98
  with smooth errors and 100 curves, at about half of REML + CL2's
  width) and for Gaussian and Poisson responses, but it covers binary
  \(\alpha(t)\) and \(\gamma(t)\) at only 0.89. Over the combinations of
  truth and residual source on the six plasmode datasets with
  independent source curves, it undercovers \(\beta\) (0.87 on average;
  NCV + CL2 0.82, REML + CL2 0.93). Residual smoothing bias of the NCV
  estimate is a plausible contributor; variance and tail errors are not
  excluded (Supplement S4.2).
\item
  \emph{A smoothing-parameter uncertainty term for REML + CL2} raises
  coverage of \(\beta\) by 1.2--1.6 pp without improving the interval
  score (Supplement S4.3).
\item
  \emph{An AR(1) working model} (\texttt{bam()} with AR(1) errors
  restarting at each curve, correlation profiled by the
  restricted-likelihood criterion; Appendix B.1) covers 0.44--0.93 on
  the nine plasmode datasets (Supplement S3.1).
\item
  \emph{The curve bootstrap} (resampling curves and refitting) covers
  0.93--0.97 on real residual curves, with intervals 1.4 times as wide
  as REML + CL2's and a median of 507 s per analyzed dataset (Supplement
  S3.2).
\item
  \emph{Penalized functional GEE}
  (\citeproc{ref-LoewingerLevisCuiPereira2025}{Loewinger et al. 2026}),
  in the R package fastFGEE (version 0.2.2). Under dependent errors, it
  improves the surface over REML by less than NCV (mean squared error
  1.3--3.7 times NCV's). This comparison combines the one-step update
  with the difference between REML and NCV tuning of the pffr fit it
  starts from. Simulations by G. Loewinger (personal communication)
  separate the two by initializing the update from the NCV fit, with our
  functional error correlation, AR(1) correlation between the curves of
  a subject and 40 subjects: with one curve per subject, as in our
  comparison, the update had 1.48 times the mean squared error of the
  NCV fit for \(\beta(s,t)\); with 5, 25 and 100 curves per subject,
  0.67, 0.54 and 0.50 times (Supplement S3.3). With repeated curves per
  subject, its longitudinal working correlation thus improves on
  curve-blocked NCV. Wald intervals from its sandwich covariance cover
  \(\beta(s,t)\) at only 0.67--0.89: their standard errors are well
  below the spread of the estimate (Supplement S3.3). Its interface
  offers a wild-bootstrap critical value for intercepts and scalar
  effects; we obtained it for the surface and fitted means from an
  internal routine, which is not a documented use of the package. With
  an independence working correlation and its \(t\) adjustment, these
  intervals for \(\beta(s,t)\) were conservative under dependence in the
  simulation and slightly narrower than REML + CL2 with the
  Satterthwaite critical value (coverage 0.97--0.99 against 0.95--0.96,
  lower tail 0.94--0.98 against 0.92--0.94, at 0.88--0.99 times the
  width). On the 6 plasmode datasets with independent source curves
  (midpoint truth, REML residuals) and all subjects, they were again
  narrower (0.7--0.9 times) and covered 0.93--0.98 against 0.94--0.96,
  but on 5 of them their coverage was much less even over the grid
  (lower tail 0.75--0.85 against 0.89--0.92). The bootstrap calibrates
  one critical value for the whole surface, so it targets average
  coverage. With 40 subjects, the \(t\) adjustment often reached its
  floor of one degree of freedom and made the intervals 1.5--4.9 times
  as wide as those of REML + CL2 with the Satterthwaite critical value;
  without it, they covered as low as 0.89, against 0.94--0.96. Results
  for fitted means and for independent errors are in Supplement S3.3.
\item
  \emph{Clusters other than curves.} Block- and subject-clustered CL2
  covered below nominal on the weather data, with 16 spatial blocks as
  clusters, and in a function-on-scalar design with 25 subjects and five
  curves per subject. With 100 or 500 subjects, subject-clustered CL2
  was close to nominal in that design (Supplement S2.3, S3.3).
\item
  \emph{A principal-component term for the covariate.} With the low-rank
  covariate of one synthetic setting, an \texttt{ffpc()} term with the 3
  components that explain 95\% of the covariate's variance estimated
  \(\beta(s,t)\) about as well as NCV with \texttt{ff()} and much better
  than REML with \texttt{ff()}; its CL2 intervals with the normal
  critical value covered 0.91 (0.95 with five components; Supplement
  S1.9).
\end{itemize}

\section{Plasmode study}\label{sec-plasmode}

\subsection{Design}\label{sec-plasmode-design}

A plasmode study simulates from real data with a known truth
(\citeproc{ref-FranklinSchneeweissPolinskiRassen2014}{Franklin et al.
2014}). Each of nine datasets has Gaussian responses and is fitted with
the model in Equation~\ref{eq-model} by REML and by curve-blocked NCV. A
fit's fitted means serve as the truth: the \emph{REML-fitted truth}, the
\emph{NCV-fitted truth}, and a \emph{midpoint truth} refitted with every
smoothing parameter at the geometric mean of the two fits' values. A
replicate adds each subject's real residual curve, from the REML or the
NCV fit, with a random sign (a \emph{curve flip}) to the truth at its
own covariates. This keeps the covariates, the residual shapes and their
within-curve dependence, and makes the truth known. Because the signs
are independent, the simulated errors are independent across curves, and
each subject's error takes only two values, plus or minus its residual
curve: the construction is a Rademacher wild cluster bootstrap around
the fitted residuals, not a model of new real curves. Its truths are
exactly representable by the fitted model, so it cannot evaluate
misspecification of the mean. Two further limits follow from the
construction. A curve's contribution to the sandwich varies little
between replicates, so the plasmode study tests whether CL2's standard
error matches the spread of the estimate, but hardly the Satterthwaite
reference distribution, which accounts for the variability of that
standard error; the evidence for the reference distribution is the
simulation. Fitted residuals are also somewhat smaller than the errors
that produced them, more so for a more flexible fit. A further error
source for the ECG strain data, heartbeat difference curves, is not
shaped by a fit (Appendix B.3).

In six datasets, the source curves are independent: ECG-derived
myocardial strain, atrial strain from a randomized trial, knee moments
during running and walking, the DTI tract profiles and an eight-lead ECG
dataset. In three, the source curves are serially or spatially dependent
(monthly ocean profiles, weather stations, weekly electricity demand);
their simulated errors are still independent across curves, except in
block-flip runs reported in the supplement. The midpoint fit lies
between the REML and NCV fits for every term on only 3 of the 9
datasets, and its surface is rougher than both on running, ocean,
weather; it is a third smoothing level, not a compromise. The datasets
have 73--138 curves on 30--101 grid points, design effects of the
residual curves (the factor by which within-curve correlation inflates
the variance of a grid mean) of 6--40 and functional covariates of rank
4--24 (Appendix B.2). Each combination of dataset, truth and residual
source is run with all subjects and with 40, with 200 replicates; the
mean is scored for a fixed set of 50 subjects.

\subsection{Results}\label{sec-plasmode-results}

\begin{figure}

\centering{

\pandocbounded{\includegraphics[keepaspectratio]{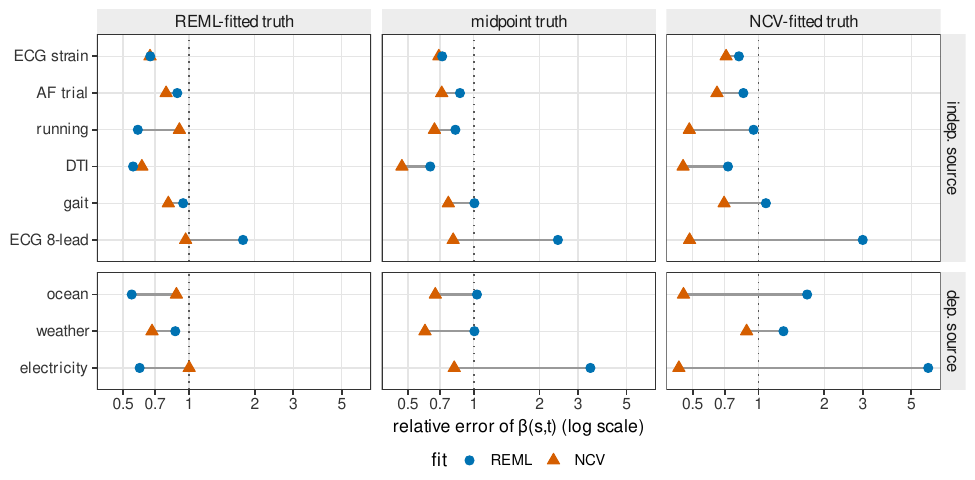}}

}

\caption{\label{fig-plasmode-est}Plasmode relative error of the REML and
NCV estimates of \(\beta(s,t)\), all subjects, by dataset and truth
source (median over replicates and the two residual sources; log scale;
dotted: error as large as the effect).}

\end{figure}%

\begin{figure}

\centering{

\pandocbounded{\includegraphics[keepaspectratio]{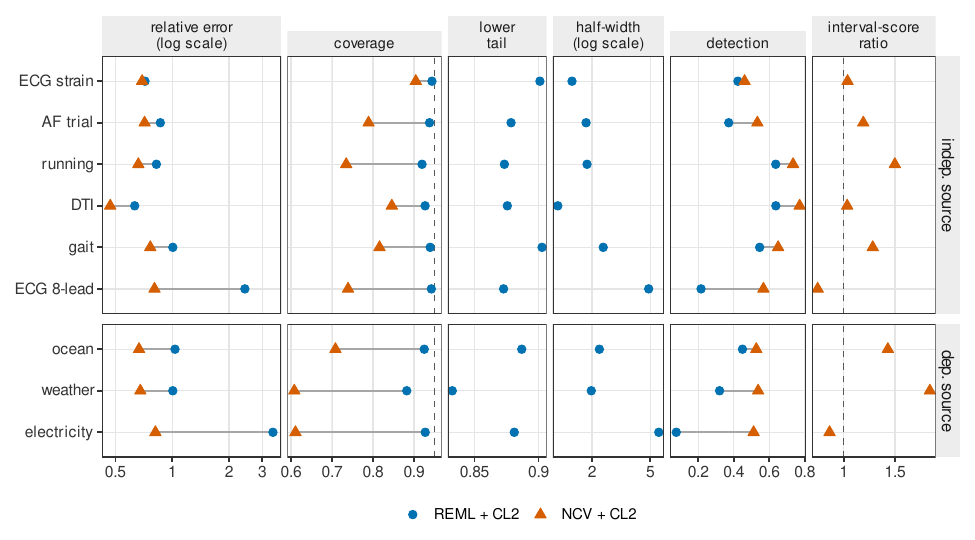}}

}

\caption{\label{fig-plasmode-info}Plasmode study, \(\beta(s,t)\), all
subjects, per dataset: median relative error of the REML and NCV
estimates; coverage, detection rate; for REML + CL2 the lower tail (5\%
quantile of pointwise coverage) and the relative half-width; for NCV +
CL2 the interval score relative to REML + CL2 (means, medians or
geometric means over truth and residual sources). Gray segments join the
two intervals of a dataset; log scales for relative error and
half-width. Dashed: 0.95 for coverage, 1 for the ratio. Values:
Supplement S7.}

\end{figure}

\emph{Estimation.} The median relative errors are 0.6--3.4 for REML and
0.5--0.8 for NCV across datasets (Figure~\ref{fig-plasmode-est},
Figure~\ref{fig-plasmode-info}). Weak identification of \(\beta\) in
directions the functional covariate barely excites is a possible reason
for their size (Section~\ref{sec-intro}). Which fit is better depends on
the truth: NCV's relative error is 0.07--0.87 times REML's under the
NCV-fitted truth, 0.24--0.97 times under the midpoint truth and
0.55--1.69 times under the REML-fitted truth, where NCV is worse on 4 of
the 9 datasets. A fit used as truth is shaped like the selector that
produced it, so the truth source favors that selector; the simulation,
whose truths come from neither fit, is the cleaner comparison, although
part of its gap reflects its smooth truths
(Section~\ref{sec-res-estimation}). For the mean, NCV's relative error
is 0.71--1.00 times REML's per dataset.

\begin{figure}

\centering{

\pandocbounded{\includegraphics[keepaspectratio]{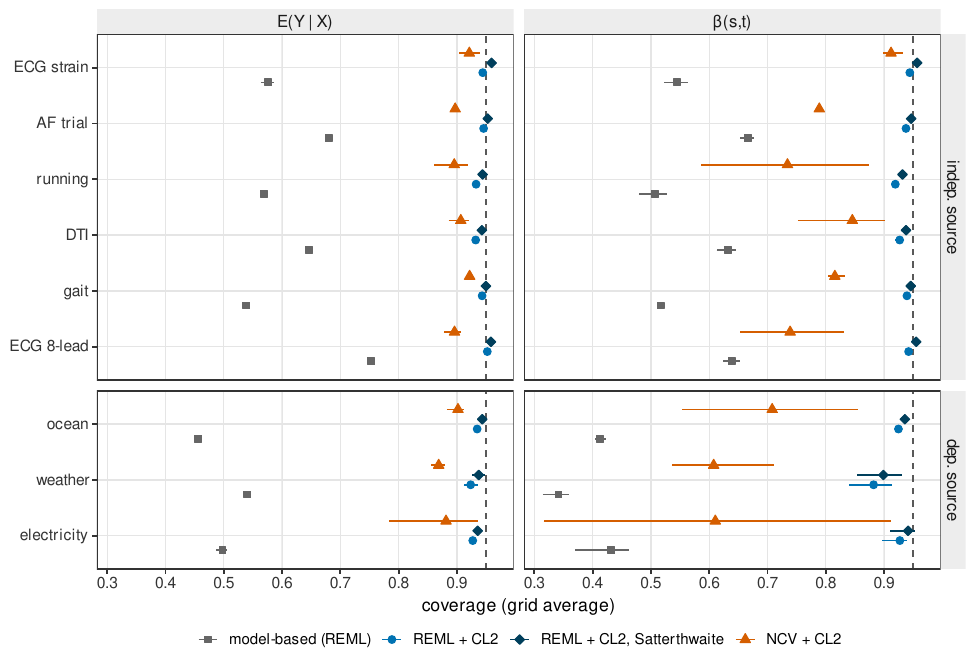}}

}

\caption{\label{fig-plasmode}Plasmode coverage of nominal 95\% pointwise
intervals for the mean and \(\beta(s,t)\), all subjects: mean over truth
and residual sources (symbols) and their range (bars). REML + CL2 and
NCV + CL2 use the normal critical value. Dashed: 0.95.}

\end{figure}%

\emph{Intervals.} With curve flips, the simulated errors are independent
across curves on all nine datasets, so CL2's assumption holds on all of
them. Model-based intervals cover 0.34--0.75 for the mean and \(\beta\).
REML + CL2 covers the mean at 0.92--0.95 on all nine datasets and
\(\beta\) at 0.92--0.94 on eight (Figure~\ref{fig-plasmode}), but
\(\beta\) at only 0.88 on the weather data, which combine few curves
(73), a functional covariate of rank 4 and rough residuals.

With the normal critical value, REML + CL2's coverage of \(\beta\) is
below 0.95 in 130 of 130 combinations of dataset, truth, residual
source, number of subjects and construction variant. Among the 120
combinations with curve flips, it is below 0.95 in 120 with the normal
and in 63 with the Satterthwaite critical value. These combinations
share data and are not independent confirmations. Its standard error is
somewhat smaller than the replicate spread of the estimate (median ratio
0.89--1.01 per dataset), which is consistent with a standard error a
little short, plus further error on some datasets. With 40 subjects on
the six datasets with independent source curves, REML + CL2 covers
\(\beta\) 1.4--4.8 pp below nominal and the mean 0.6--3.2 pp below; on
the weather data, it covers \(\beta\) at 0.77--0.87.

With Satterthwaite degrees of freedom (Supplement S2.1), REML + CL2
covers \(\beta\) at 0.94 on average over the nine datasets with all
subjects and 0.94 with 40 (normal critical value: 0.93 and 0.91), at
1.05 and 1.13 times the width. It covers the mean, \(\alpha(t)\) and
\(\gamma(t)\) at 0.94--0.96. The weather surface stays short (0.90 with
all subjects, 0.86 with 40). Adding the smoothing-parameter uncertainty
term and using the \(t_{G-1}\) critical value helps about as much, with
the weather surface again short (Supplement S4.3).

On the six datasets with independent source curves, NCV + CL2 covers
\(\beta\) at 0.73--0.91, with lower tails 0.44--0.81, and loses up to 19
pp with 40 subjects. Its standard error is short (median ratio to the
replicate standard deviation 0.75--0.98 across these datasets). A refit
with the smoothing parameters fixed at their median restores it
(1.03--1.15), but coverage then still stays at 0.69--0.92. Fixing the
smoothing parameters changes both the amount and the variability of
smoothing, so this is a diagnostic, not an exact decomposition; it
suggests smoothing bias as the larger part (Supplement S4.1).

\emph{Informativeness.} As in the simulation, REML + CL2 is close to
nominal but wide for \(\beta\): its half-width is 1.2--5.7 times the
effect, and it detects 0.08--0.64 of the grid points where the true
value exceeds 25\% of its maximum. NCV + CL2 detects more (0.46--0.77)
but does not hold its level, and here its interval score is 0.7--1.8
times REML + CL2's (higher on 7 of the 9 datasets).

Intervals centered at the NCV estimate with a bias correction and a
smoothing-parameter term (Section~\ref{sec-res-other}) cover \(\beta\)
at 0.83--0.94 on the six datasets with independent source curves,
against 0.91--0.94 for REML + CL2, at 0.7--1.1 times its width and with
lower tails as low as 0.38 in single combinations. On the three datasets
with dependent source curves, they cover 0.66--0.84 (Supplement S4.1).
These comparisons are descriptive: the NCV-centered intervals were
computed on 50 replicates and REML + CL2 on 100, without a paired
measure of precision.

The NCV estimate with REML + CL2's half-width covers 0.98--1.00 in the
simulation (dependent errors, 100 curves, smooth and rough truths) and
0.97 and 0.96 under the NCV-fitted and the midpoint plasmode truth (all
subjects, mean over datasets). Under the REML-fitted truth, it covers
0.91 with a lower tail of 0.67, against a lower tail of 0.87 for REML +
CL2. It is no narrower than REML + CL2 and detects less on all nine
datasets (Supplement S4.1).

\section{Recommendations and software}\label{sec-recommend}

\begin{tcolorbox}[enhanced jigsaw, colbacktitle=quarto-callout-note-color!10!white, titlerule=0mm, colback=white, arc=.35mm, left=2mm, opacityback=0, toprule=.15mm, leftrule=.75mm, coltitle=black, bottomtitle=1mm, title=\textcolor{quarto-callout-note-color}{\faInfo}\hspace{0.5em}{Recommendations for applied work with pffr()}, toptitle=1mm, rightrule=.15mm, bottomrule=.15mm, breakable, opacitybacktitle=0.6, colframe=quarto-callout-note-color-frame]

\begin{enumerate}
\def\labelenumi{\arabic{enumi}.}
\tightlist
\item
  \textbf{Assume within-curve dependence.} Do not use model-based
  intervals of the working-independence fit, and expect REML to
  undersmooth coefficient surfaces.
\item
  \textbf{Shape of a surface: curve-blocked NCV.} Fit the model with
  \texttt{method\ =\ "NCV"} and show its surface estimate as the
  description of the effect.
\item
  \textbf{Inference: REML + CL2 with Satterthwaite degrees of freedom},
  for all terms and for fitted means. The intervals for a surface are
  wide; report them as they are, and as pointwise intervals.
\item
  \textbf{Both goals: fit twice.} Take the surface from the NCV fit and
  the intervals from the REML fit, and show them side by side, not as a
  band around the NCV surface.
\item
  \textbf{Means, intercepts and scalar effects only:} one REML fit with
  these intervals is enough, with the exceptions in 6.
\item
  \textbf{Treat as approximate:} intervals for intercepts and scalar
  effects of binary responses and for misregistered counts, and any
  interval with fewer than 20 curves or with few clusters of curves.
\item
  \textbf{Check the basis.} Refit with a richer basis and compare; do
  not interpret the shape of a surface that changes with the basis.
\end{enumerate}

\end{tcolorbox}

The evidence for each recommendation and its limits are as follows.

\emph{Dependence (1).} Under dependence, REML undersmooths coefficient
surfaces (Section~\ref{sec-res-estimation}), and model-based intervals
undercover, more so on dense grids (Section~\ref{sec-res-intervals}).

\emph{Shape of a surface (2).} The forthcoming refund releases set one
neighborhood per curve (per \texttt{cluster} if given) for
\texttt{method\ =\ "NCV"}; in earlier releases and in plain
\texttt{gam()}, \texttt{method\ =\ "NCV"} alone leaves out single
points, and the neighborhoods must be supplied through \texttt{nei}.
NCV's gain in estimation also holds, more weakly, for smooth effects of
scalar covariates that vary along \(t\)
(Section~\ref{sec-res-estimation}). When the surface was zero, the NCV
estimate stayed much closer to zero than REML's, which showed structure
(Section~\ref{sec-res-null}). NCV is not uniformly better: it was worse
on some plasmode datasets under the REML-fitted truth, which favors REML
(Section~\ref{sec-plasmode-results}). Apart from misregistration, a
small share of its fits under dependence have an unusually small
smoothing parameter and a larger error, which convergence flags rarely
reveal outside scaled-\(t\) responses, although NCV still estimated the
surface better than REML on nearly all of these replicates
(Section~\ref{sec-res-estimation}). For Gaussian responses, pcre and GLS
with an FPCA residual covariance also estimated the surface much better
than REML, but worse than NCV in the simulation and comparably to it in
a plasmode comparison (Section~\ref{sec-res-estimation}). With one curve
per subject, as in our fastFGEE comparison, fastFGEE's one-step update
did not improve on NCV, whether started from the REML or the NCV fit;
with repeated curves per subject, the update started from the NCV fit
estimated the surface better than NCV in simulations by G. Loewinger
(Section~\ref{sec-res-other}), so for such data it is worth considering
for the shape of a surface. If mgcv flags an NCV fit as not converged,
as it usually did for scaled-\(t\) responses, refitting from other
starting smoothing parameters and comparing the criterion and the
surface is a simple check (Appendix B.1); it was not evaluated as a
procedure.

\emph{Inference (3).} The recommended interval uses the curve-clustered
CL2 covariance of the REML fit with the full block and the Satterthwaite
critical value. It was evaluated for all four estimands in the main
factorial at the default signal with 20, 40 and 100 curves, in all
extensions and in the plasmode study. Under dependence, it was close to
nominal for the mean, \(\beta(s,t)\) and \(\gamma(t)\) in the main
factorial and for \(\beta(s,t)\) on average in the plasmode study, but
not on one dataset with few curves and a low-rank covariate
(Section~\ref{sec-res-intervals}, Section~\ref{sec-plasmode-results}).
Under independent errors, it was conservative for \(\beta(s,t)\)
(Section~\ref{sec-res-intervals}). The price is width, which is largest
with few curves. fastFGEE's wild-bootstrap intervals for \(\beta(s,t)\),
obtained from an internal routine, were conservative and slightly
narrower under dependence in the simulation, but we keep REML + CL2
because on most of the six plasmode datasets with independent source
curves its pointwise coverage was more even, and it kept its coverage at
40 subjects without the large widening of fastFGEE's \(t\) adjustment
(Section~\ref{sec-res-other}). Only pointwise intervals were evaluated:
a pointwise band answers ``is there an effect, and where'' poorly
(Section~\ref{sec-plasmode-results}) and ``is \(\beta(s,t)\) zero
everywhere'' not at all, which needs simultaneous bands or global tests
(Section~\ref{sec-res-null}).

\emph{Fitting twice (4, 5).} The two fits differ only in how the
smoothing parameters are chosen; Section~\ref{sec-res-tradeoff} explains
why they differ. In our Gaussian benchmark with 100 curves, an NCV fit
took about 4 times as long as a REML fit (Appendix C). Disregard the NCV
fit's standard errors: NCV + CL2 intervals undercover
(Section~\ref{sec-res-tradeoff}, Section~\ref{sec-plasmode-results}).
One way to report both fits is the NCV surface next to REML + CL2's map
of grid points whose interval excludes zero, labeled as pointwise. Under
a zero surface, most replicates show some excluded points
(Section~\ref{sec-res-null}); this display and disagreements between the
two fits were not evaluated. In the dependent main-factorial settings at
the default signal, the NCV estimate lay outside the Satterthwaite
interval of REML + CL2 at 1.2--2.4\% of the grid points. Centering REML
+ CL2's half-width at the NCV estimate is not a shortcut to a band
around the NCV surface: it overcovers in the simulation, covers unevenly
under the REML-fitted plasmode truth, and detects less than REML + CL2
(Section~\ref{sec-plasmode-results}). For the functional intercept,
scalar effects and fitted means, REML + CL2 gives both the estimate and
the intervals, with the exceptions that follow.

\emph{Approximate intervals (6).} Intervals for the intercept and scalar
effects of binary responses and for misregistered counts undercover,
also with Satterthwaite degrees of freedom
(Section~\ref{sec-res-intervals}). With 20 curves and dependent errors
in the main factorial, the Satterthwaite critical value brought the
mean, \(\beta(s,t)\) and \(\gamma(t)\) close to nominal, but not binary
intercepts; with 20 curves and independent errors, every interval
undercovered binary intercepts and scalar effects
(Section~\ref{sec-res-intervals}). Nothing below 20 curves, or with 20
curves outside the main factorial, was evaluated; treat intervals there
as approximate. With several curves per subject, cluster by subject
(\texttt{cluster\ =}). Subject clustering leaves fewer clusters and does
not address dependence between subjects; block- and subject-clustered
CL2 covered below nominal with 16 blocks and with 25 subjects, and close
to nominal with 100 or 500 subjects (Section~\ref{sec-res-other}). With
\texttt{method\ =\ "NCV"}, the same argument gives subject
neighborhoods, which were not evaluated.

\emph{Basis and identification (7).} A richer basis protects against
approximation bias but widens REML + CL2's intervals
(Section~\ref{sec-res-tradeoff}). It does not help where the functional
covariate carries little information about a direction of the surface:
the surface is then weakly identified
(\citeproc{ref-ScheiplGreven2016}{Scheipl and Greven 2016}). Refrain
from interpreting the surface's shape when the estimate or its interval
changes substantially with the basis, or when the covariate's principal
components leave the relevant directions poorly covered given the
penalty's null space; agreement between two penalized fits does not
certify identification. Fitted means or well-supported contrasts are
then safer targets, though not protected against misspecification
(Supplement S5). If the functional covariate's variation is concentrated
in a few principal components, an \texttt{ffpc()} term with a small
number of components can replace \texttt{ff()}
(Section~\ref{sec-res-other}). For covariates with a richer spectrum,
keep \texttt{ff()}: few components truncate the surface and the
intervals undercover, and a high variance-explained threshold, such as
0.99 (the default of current refund releases), gives unstable, very
noisy estimates, as is known for functional principal component
regression, an ill-posed problem (\citeproc{ref-HallHorowitz2007}{Hall
and Horowitz 2007}; \citeproc{ref-ReissOgden2007}{Reiss and Ogden
2007}).

\emph{Software.} The study used refund at development commit 79a346fb,
which contains these procedures, and an unpatched and a patched
pre-release build of mgcv 1.9-5; the patch holds the corrections found
in this study, among them those of mgcv's covariances for NCV fits
(Supplement S4.2, S6). Forthcoming releases of refund and mgcv will
implement these recommendations, with REML + CL2 and Satterthwaite
degrees of freedom as the default intervals of \texttt{pffr()} fits.
Supplement S6 gives example code, the software used in the study and a
reproduction route.

\section{Discussion}\label{sec-discussion}

Under within-curve dependence, the working-independence REML fit of a
penalized function-on-function regression gives an undersmoothed, noisy
estimate of the coefficient surface and model-based intervals that are
far too narrow. Curve-blocked NCV, which typically smooths more,
improves the estimate but not the intervals. A curve-clustered sandwich
with a leverage adjustment for the penalized hat and Satterthwaite
degrees of freedom brings the intervals close to nominal for most
estimands of Gaussian, count and binary responses in the simulation,
down to 20 curves, and for the surface on eight of the nine plasmode
datasets, but it cannot improve the estimate it is centered on. The two
remedies thus show the undersmoothing-versus-bias-correction tension of
nonparametric inference (\citeproc{ref-Hall1992}{Hall 1992};
\citeproc{ref-CalonicoCattaneoFarrell2018}{Calonico, Cattaneo, and
Farrell 2018}), here produced by the dependence, hence our
recommendation to fit twice when the surface matters. Intermediate
amounts of smoothing between REML and NCV, global tests of
\(\beta \equiv 0\) and simultaneous bands are the natural next steps, as
is a wild cluster bootstrap critical value
(\citeproc{ref-CameronGelbachMiller2008}{Cameron, Gelbach, and Miller
2008}) for REML + CL2 with few curves; they were not studied here. The
evidence has the following limits.

\begin{itemize}
\tightlist
\item
  \emph{Binary responses.} REML + CL2 undercovers intercepts and scalar
  effects of binary responses, also under independent errors and with
  Satterthwaite degrees of freedom. The diagnostics are compatible with
  a standard error that is too small for these terms, partly through
  smoothing-parameter variability; local bias is not excluded. Every
  interval centered at the NCV estimate undercovers for binary responses
  (Supplement S4.1).
\item
  \emph{Misregistration.} When curves are observed on their own clocks,
  the observed-clock mean of a count response is not a generalized
  linear predictor, and no method covers it; for Gaussian responses,
  REML + CL2 is unaffected.
\item
  \emph{Design.} The synthetic function-on-function design has 20 curves
  only in the main factorial and at most 100 curves, a zero surface only
  for Gaussian and binary responses with independent and smooth errors,
  one empirically derived smooth error kernel (from the DTI data) beside
  parametric kernels, no dependent error process of intermediate
  strength, regular grids without missing values, and first-order
  difference penalties throughout. For count and binary responses,
  dependence enters only through a Gaussian copula.
\item
  \emph{Mean model.} The study addresses misspecified dependence under a
  correct model for the mean, and all synthetic truths in the main study
  are projected onto the fitted bases. For the smooth \(\beta(s,t)\) of
  the Gaussian default setting, the projection is immaterial: scored
  against the exact function, REML + CL2 covers it at 0.93--0.94 under
  smooth errors (Supplement S5.1, normal critical value). A basis that
  cannot represent the effect is a misspecified mean model, and none of
  the intervals studied here protects against it (Supplement S5.1). With
  a local bump of standard deviation 0.05 added to the surface, REML +
  CL2 covers \(\beta(s,t)\) at only 0.17--0.58 inside the bump. For a
  truth with components the covariates barely excite and the default
  basis cannot represent, it covers \(\beta(s,t)\) at only 0.44--0.56
  under smooth errors and 0.30 under independent errors. A basis with
  four times finer knot spacing restores 0.93 under smooth errors, at
  the price of wider intervals, but only 0.86 under independent errors;
  NCV + CL2 stays short even then. These coverages refer to the exact
  truth, which mixes approximation bias with weak identification.
\item
  \emph{Plasmode study.} Fits used as truths favor their own selector,
  the errors are sign-flipped fitted residuals with independent signs
  across curves, the truths are representable by the model, and all nine
  datasets are Gaussian and were chosen, not sampled. With sign flips,
  the plasmode results test CL2's standard error, but hardly the
  Satterthwaite reference distribution
  (Section~\ref{sec-plasmode-design}).
\item
  \emph{Intervals around the NCV estimate.} Their shortfall is mainly
  smoothing bias in the simulation; on the plasmode data, the
  diagnostics suggest smoothing bias and the variability of the selected
  smoothing parameters, but fixing the smoothing parameters changes
  both. Neither a one-step bias correction nor a variance term for the
  smoothing parameters, nor their combination as a frozen-map
  approximation, gave a narrower calibrated interval in the plasmode
  study. Calibrated intervals after data-driven heavier smoothing remain
  open for the NCV fit. A wild-bootstrap critical value around
  fastFGEE's fit gave conservative intervals for \(\beta\), narrower
  than REML + CL2 with the Satterthwaite critical value, under
  dependence in the simulation. On most of the plasmode datasets with
  independent source curves, its coverage was less even with all
  subjects, and with 40 subjects its intervals were very wide with the
  \(t\) adjustment and undercovered without it. Started from the REML
  fit, its mean squared error for \(\beta\) under dependence in the
  simulation is still 1.3--3.7 times NCV's; started from the NCV fit, it
  improves on NCV with repeated curves per subject, but not with one
  curve per subject as in our comparison (Section~\ref{sec-res-other}).
  Its intervals with repeated curves per subject were not evaluated
  here.
\end{itemize}

\section*{Code and data availability}\label{code-and-data-availability}
\addcontentsline{toc}{section}{Code and data availability}

Code and summary files are at \url{https://github.com/fabian-s/fof-inf};
the tables and figures of this paper and its supplement are computed
from those summary files. The per-replicate records are at
\url{https://doi.org/10.5281/zenodo.23916388}. Records of the two
cardiology datasets are patient and volunteer data and are not shared;
only aggregate results are available for them. The other datasets are
public (Appendix B.2).

\section*{Acknowledgements}\label{acknowledgements}
\addcontentsline{toc}{section}{Acknowledgements}

I thank Gabriel Loewinger for discussions of functional GEE and
fastFGEE, and for sharing the simulation results cited in Supplement
S3.3.

\section*{Use of generative AI}\label{use-of-generative-ai}
\addcontentsline{toc}{section}{Use of generative AI}

All code and all text of this paper and its supplement were generated by
large language models: Claude Opus 5.5 and Claude Fable 5.1 (Anthropic)
and ChatGPT models Sol and Astra, versions 5.6 and 6 (OpenAI). The study
design, structural decisions and research directions were developed
collaboratively between the author and these models. The author takes
full responsibility for the content.

\section*{References}\label{references}
\addcontentsline{toc}{section}{References}

\phantomsection\label{refs}

\appendix

\section{The CL2 covariance}\label{sec-app-cl2}

Notation follows the main text: \(\mathbf W\) holds the working weights
\(w_{ij} = (d\mu/d\eta)^2_{ij}/\{\phi V(\mu_{ij})\}\),
\(\mathbf S = \sum_r (\lambda_r/\phi)\mathbf S_r\),
\(\mathbf V_p = (\mathbf X^\top\mathbf W\mathbf X + \mathbf S)^{-1}\),
\(\mathbf V_e = \mathbf V_p\mathbf X^\top\mathbf W\mathbf X\mathbf V_p\),
and for curve \(i\)
\(\tilde{\mathbf X}_i = \mathbf W_i^{1/2}\mathbf X_i\) and
\(z_{ij} = (y_{ij} - \hat\mu_{ij})/\{\phi V(\hat\mu_{ij})\}^{1/2}\). The
stacked whitened design is \(\tilde{\mathbf X}\), and
\(\tilde{\mathbf H} = \tilde{\mathbf X}\mathbf V_p\tilde{\mathbf X}^\top\)
has blocks \(\tilde{\mathbf H}_{ih}\).

\subsection{Scores and the penalized
bread}\label{scores-and-the-penalized-bread}

For an exponential-family response, the score of the log-likelihood
contribution of \(y_{ij}\) is
\begin{equation}\phantomsection\label{eq-s1-score}{
\mathbf s_{ij}(\boldsymbol\theta)
= \frac{(y_{ij} - \mu_{ij})\,(d\mu/d\eta)_{ij}}{\phi V(\mu_{ij})}\,\mathbf x_{ij}
= z_{ij}\,\tilde{\mathbf x}_{ij},
\qquad \tilde{\mathbf x}_{ij} = w_{ij}^{1/2}\mathbf x_{ij},
}\end{equation} so the score of curve \(i\) is
\(\mathbf u_i = \sum_j \mathbf s_{ij} = \tilde{\mathbf X}_i^\top\mathbf z_i\).
The fit maximizes
\(\sum_{i,j}\ell_{ij}(\boldsymbol\theta) - \tfrac12\boldsymbol\theta^\top\mathbf S\boldsymbol\theta\),
whose estimating equation is
\begin{equation}\phantomsection\label{eq-s1-ee}{
\sum_{i=1}^G \mathbf u_i(\hat{\boldsymbol\theta}) = \mathbf S\hat{\boldsymbol\theta}.
}\end{equation} The negative Hessian of the penalized log-likelihood is
\(\mathbf X^\top\mathbf W\mathbf X + \mathbf S = \mathbf V_p^{-1}\)
(exactly for canonical links, in expectation otherwise). Linearizing
Equation~\ref{eq-s1-ee} around the target \(\boldsymbol\theta^*\) at
which its expectation vanishes gives
\begin{equation}\phantomsection\label{eq-s1-lin}{
\hat{\boldsymbol\theta} - \boldsymbol\theta^* \approx \mathbf V_p \sum_{i=1}^G \{\mathbf u_i(\boldsymbol\theta^*) - \mathrm E\,\mathbf u_i(\boldsymbol\theta^*)\},
\qquad
\mathrm{Var}(\hat{\boldsymbol\theta}) \approx \mathbf V_p \Bigl\{\sum_{i=1}^G \mathrm{Var}\,\mathbf u_i(\boldsymbol\theta^*)\Bigr\}\mathbf V_p ,
}\end{equation} so \(\mathbf V_p\) is the bread and only the meat
\(\sum_i\mathrm{Var}\,\mathbf u_i\) depends on the within-curve
dependence. Under the working model,
\(\mathrm{Var}(\mathbf z_i) = \mathbf I\) and the meat is
\(\tilde{\mathbf X}^\top\tilde{\mathbf X}\), which returns
\(\mathbf V_e\). The unpenalized bread
\((\mathbf X^\top\mathbf W\mathbf X)^{-1}\) linearizes a different
estimator. Scores and penalty are both on the scale of the summed
log-likelihood; a construction that averages the scores over the \(N\)
observations must use the penalty
\(\sum_r\lambda_r\mathbf S_r/(N\phi)\), or the bread and the shrinkage
are mis-scaled by the factor \(N\) (and by \(\phi\) if it is dropped).

By Equation~\ref{eq-s1-ee}, the curve scores at the estimate do not sum
to zero:
\(\sum_i \tilde{\mathbf X}_i^\top\mathbf z_i = \mathbf S\hat{\boldsymbol\theta}\).
Unlike the unpenalized cluster sandwich
(\citeproc{ref-LiangZeger1986}{Liang and Zeger 1986}), whose curve
scores at the estimate sum to zero, the meat
\(\sum_i\mathbf U_i\mathbf U_i^\top\) below is not centered; Proposition
A.1(b) gives the consequence.

\subsection{The leverage adjustment}\label{the-leverage-adjustment}

The CL2 covariance of the main text is
\begin{equation}\phantomsection\label{eq-s1-cl2}{
\hat{\mathbf V}_{\mathrm{CL2}} = \frac{G}{G-1}\,\mathbf V_p\Bigl(\sum_{i=1}^G\mathbf U_i\mathbf U_i^\top\Bigr)\mathbf V_p + (\mathbf V_p - \mathbf V_e),
\qquad
\mathbf U_i = \tilde{\mathbf X}_i^\top\mathbf R_i^{-1/2}\mathbf z_i ,
}\end{equation} with
\(\mathbf R_i = [(\mathbf I - \tilde{\mathbf H})(\mathbf I - \tilde{\mathbf H})^\top]_{ii}\).
Since
\(\tilde{\mathbf X}^\top\tilde{\mathbf X} = \mathbf V_p^{-1} - \mathbf S\),
\((\mathbf I - \tilde{\mathbf H})^2 = \mathbf I - \tilde{\mathbf X}(2\mathbf V_p - \mathbf V_p\tilde{\mathbf X}^\top\tilde{\mathbf X}\mathbf V_p)\tilde{\mathbf X}^\top\)
and \begin{equation}\phantomsection\label{eq-s1-Rg}{
\mathbf R_i
= \mathbf I - \tilde{\mathbf X}_i\bigl(2\mathbf V_p - \mathbf V_p\tilde{\mathbf X}^\top\tilde{\mathbf X}\mathbf V_p\bigr)\tilde{\mathbf X}_i^\top
= \mathbf I - \tilde{\mathbf H}_{ii} - \tilde{\mathbf X}_i\mathbf V_p\mathbf S\mathbf V_p\tilde{\mathbf X}_i^\top .
}\end{equation} \(\mathbf R_i\) needs only \(\tilde{\mathbf X}_i\) and
one \(p\times p\) matrix. It is computed in the column space of
\(\tilde{\mathbf X}_i\) in \(\mathbb R^{D_i}\) (dimension at most
\(\min(D_i, p)\)), because \(\mathbf R_i\) equals \(\mathbf I\) on its
orthogonal complement and \(\tilde{\mathbf X}_i^\top\) annihilates that
complement.

\emph{Relation to CR2.} The bias-reduced linearization of Bell and
McCaffrey (\citeproc{ref-BellMcCaffrey2002}{2002}) chooses
\(\mathbf A_i\) with
\(\mathbf A_i[(\mathbf I - \mathbf H)\boldsymbol\Phi(\mathbf I - \mathbf H)^\top]_{ii}\mathbf A_i = \boldsymbol\Phi_i\)
for a working covariance \(\boldsymbol\Phi\)
(\citeproc{ref-McCaffreyBell2006}{McCaffrey and Bell 2006};
\citeproc{ref-PustejovskyTipton2018}{Pustejovsky and Tipton 2018}). With
\(\boldsymbol\Phi = \mathbf I\) this is
\(\mathbf A_i = \mathbf R_i^{-1/2}\). For a projection,
\(\mathbf R_i = \mathbf I - \mathbf H_{ii}\) and the usual CR2 results.
The penalized hat is not idempotent: the last term of
Equation~\ref{eq-s1-Rg} is positive semidefinite, so
\(\mathbf R_i \preceq \mathbf I - \tilde{\mathbf H}_{ii}\), and the
approximation
\(\mathbf A^{\mathrm s}_i = (\mathbf I - \tilde{\mathbf H}_{ii})^{-1/2}\),
which keeps the full block \(\tilde{\mathbf H}_{ii}\) but drops the last
term of Equation~\ref{eq-s1-Rg} (the \emph{shortcut}), under-corrects:
\(\mathbf A^{\mathrm s}_i\mathbf R_i\mathbf A^{\mathrm s}_i \preceq \mathbf I\).
The dropped term is small when the per-curve leverage is small.

\emph{Expectation of the meat.} Take the Gaussian working model with
fixed \(\boldsymbol\lambda\) and fixed design:
\(\mathbf W^{1/2}\mathbf y = \tilde{\mathbf X}\boldsymbol\theta + \boldsymbol\varepsilon\)
with \(\mathrm{Var}(\boldsymbol\varepsilon) = \mathbf I\). Then
\(\mathbf z = (\mathbf I - \tilde{\mathbf H})(\tilde{\mathbf X}\boldsymbol\theta + \boldsymbol\varepsilon)\),
and
\((\mathbf I - \tilde{\mathbf H})\tilde{\mathbf X} = \tilde{\mathbf X}\mathbf V_p\mathbf S\)
gives \begin{equation}\phantomsection\label{eq-s1-mean}{
\mathrm E(\mathbf z_i) = \mathbf m_i = \tilde{\mathbf X}_i\mathbf V_p\mathbf S\boldsymbol\theta,
\qquad
\mathrm E(\mathbf z_i\mathbf z_i^\top) = \mathbf R_i + \mathbf m_i\mathbf m_i^\top .
}\end{equation} The penalized residuals have nonzero mean unless
\(\mathbf S\boldsymbol\theta = \mathbf 0\). With \(\mathbf R_i\)
invertible and the floor inactive,
\begin{equation}\phantomsection\label{eq-s1-unbiased}{
\mathrm E\Bigl(\sum_i\mathbf U_i\mathbf U_i^\top\Bigr)
= \sum_i \tilde{\mathbf X}_i^\top\tilde{\mathbf X}_i
+ \sum_i \tilde{\mathbf X}_i^\top\mathbf R_i^{-1/2}\mathbf m_i\mathbf m_i^\top\mathbf R_i^{-1/2}\tilde{\mathbf X}_i .
}\end{equation} The first term is the working-model sampling meat,
\(\mathrm{Var}(\tilde{\mathbf X}^\top\boldsymbol\varepsilon)\), so
\(\mathbf V_p(\sum_i\mathbf U_i\mathbf U_i^\top)\mathbf V_p\) is
unbiased for \(\mathbf V_e\) when the truth lies in the null space of
the penalty and biased upwards by a positive semidefinite term
otherwise. The factor \(G/(G-1)\) adds a further upward bias. Under
within-curve dependence,
\(\mathrm{Var}(\boldsymbol\varepsilon_i) = \boldsymbol\Sigma_i \neq \mathbf I\),
and the adjustment removes the residual shrinkage only approximately; it
does not whiten correlated residuals.

\emph{Floor and factor.} The eigenvalues of \(\mathbf R_i\) are floored
at \(10^{-6}\) before the inverse square root, which caps the inflation
of any residual direction at a factor \(10^3\). The floor binds only for
curves whose data nearly determine a coefficient direction on their own.
How often it bound in the REML and NCV fits of the two studies was not
recorded. In the comparator runs (Section~\ref{sec-res-tradeoff}), it
bound in no GLS fit and, among the pcre fits, only on the running data,
there in 70 of 100 replicates with all subjects and in 60 of 100 with 40
subjects, so pcre + CL2 on that dataset depends on the floor. CR2
carries no factor \(G/(G-1)\); we keep it. The factor and the uncentered
scores of Appendix A.1 are each conservative at order \(1/G\). At the
estimate, the curve scores sum to \(\mathbf S\hat{\boldsymbol\theta}\),
so the uncentered meat adds about
\(\mathbf V_p\mathbf S\hat{\boldsymbol\theta}\hat{\boldsymbol\theta}^\top\mathbf S\mathbf V_p/G\)
to the covariance, a plug-in for the squared smoothing bias divided by
\(G\). We keep both because, under dependence, the leverage adjustment
removes the residual shrinkage only approximately and CL2's standard
error is short even with them (Appendix A.6); their separate
contributions were not evaluated. The factor inflates frequentist CL2
standard errors by 1.3\% at \(G = 40\) and 0.5\% at \(G = 100\); in the
Bayesian form it multiplies only the sandwich part, so these are upper
bounds.

CR1 is Equation~\ref{eq-s1-cl2} with \(\mathbf R_i = \mathbf I\) and no
further degrees-of-freedom factor.

\subsection{Bayesian and frequentist
forms}\label{bayesian-and-frequentist-forms}

The frequentist form omits
\(\mathbf V_p - \mathbf V_e = \mathbf V_p\mathbf S\mathbf V_p\). For
Gaussian responses with identity link and fixed \(\boldsymbol\lambda\),
the bias of \(\hat{\boldsymbol\theta}\) is
\(-\mathbf V_p\mathbf S\boldsymbol\theta\). Under the smoothing prior
\(\boldsymbol\theta \sim N(\mathbf 0, \mathbf S^-)\),
\(\mathrm E(\mathbf V_p\mathbf S\boldsymbol\theta\boldsymbol\theta^\top\mathbf S\mathbf V_p) = \mathbf V_p\mathbf S\mathbf V_p\):
the allowance is the prior-averaged squared bias, and \(\mathbf V_p\)
the prior-averaged mean squared error matrix
(\citeproc{ref-Wahba1983}{Wahba 1983}; \citeproc{ref-Nychka1988}{Nychka
1988}; \citeproc{ref-MarraWood2012}{Marra and Wood 2012}). For a fixed
truth and any contrast \(\mathbf a\), by the Cauchy--Schwarz inequality,
\begin{equation}\phantomsection\label{eq-s1-cs}{
(\mathbf a^\top\mathbf V_p\mathbf S\boldsymbol\theta)^2 \le (\mathbf a^\top\mathbf V_p\mathbf S\mathbf V_p\mathbf a)\,(\boldsymbol\theta^\top\mathbf S\boldsymbol\theta),
\qquad \boldsymbol\theta^\top\mathbf S\boldsymbol\theta = \sum_r (\lambda_r/\phi)\,\boldsymbol\theta^\top\mathbf S_r\boldsymbol\theta .
}\end{equation} The allowance bounds the squared bias of every contrast
if \(\boldsymbol\theta^\top\mathbf S\boldsymbol\theta \le 1\); for
larger values the bound permits, but does not imply, a bias above the
allowance.

Where the data identify \(\boldsymbol\theta\)
(\(\tilde{\mathbf X}^\top\tilde{\mathbf X}\) positive definite), the
squared bias scales as \(\lambda^2\) and the allowance as \(\lambda\)
for small \(\boldsymbol\lambda\). Undersmoothing, as REML produces under
within-curve dependence, therefore makes the allowance large relative to
the bias but small relative to the dependence-inflated variance, and the
two forms nearly coincide. For a truth that is rough at the selected
\(\boldsymbol\lambda\)
(\(\boldsymbol\theta^\top\mathbf S\boldsymbol\theta\) large), or after
oversmoothing, the bias can exceed the allowance, most likely where the
truth is most curved. Under independent errors with a smooth truth the
allowance exceeds the bias and the Bayesian form overcovers.

\subsection{Extended families}\label{extended-families}

The CL2 construction needs a factorization
\(\mathbf s_{ij} = z_{ij}\tilde{\mathbf x}_{ij}\) in which
\(\tilde{\mathbf x}_{ij}\) carries the Fisher weight of the fit, so that
\(\operatorname{tr}\tilde{\mathbf H}\) equals the effective degrees of
freedom.

\emph{Gaussian location--scale} (mgcv's gaulss). The two linear
predictors \(\eta_1\) (mean) and \(\eta_2\) (inverse standard deviation
\(\tau\)) have block-diagonal Fisher information with weights
\(W_1 = \tau^2(d\mu/d\eta_1)^2\) and
\(W_2 = 2(d\tau/d\eta_2)^2/\tau^2\). Each observation contributes two
pseudo-rows, both assigned to its curve,
\begin{equation}\phantomsection\label{eq-s1-gaulss}{
\tilde{\mathbf x}_{k} = W_k^{1/2}\mathbf x_{k}, \qquad z_k = \frac{\partial\ell/\partial\eta_k}{W_k^{1/2}}, \qquad k = 1, 2,
}\end{equation} with
\(\partial\ell/\partial\eta_1 = \tau^2(y - \mu)\,d\mu/d\eta_1\) and
\(\partial\ell/\partial\eta_2 = (1/\tau - \tau(y-\mu)^2)\,d\tau/d\eta_2\);
\(\mathbf x_k\) is the design row of predictor \(k\), zero in the other
block.

\emph{Scaled t} (scat). With degrees of freedom \(\nu\) and scale
\(\sigma\) at their estimates, the exact location score and Fisher
weight are \begin{equation}\phantomsection\label{eq-s1-scat}{
\frac{\partial\ell}{\partial\mu} = \frac{(\nu + 1)(y - \mu)}{\nu\sigma^2 + (y - \mu)^2},
\qquad
W = \Bigl(\frac{d\mu}{d\eta}\Bigr)^2\frac{\nu + 1}{(\nu + 3)\sigma^2},
}\end{equation} and
\(z = (\partial\ell/\partial\mu)(d\mu/d\eta)/W^{1/2}\). The
working-residual score Equation~\ref{eq-s1-score} would treat the \(t\)
response as Gaussian and give no downweighting of outliers.

All other extended families of mgcv (negative binomial, Tweedie, beta,
ordered categorical) use the working-residual score
Equation~\ref{eq-s1-score} with the fitted variance function. For the
negative binomial, with its shape parameter at the estimate, this is the
exact score for the mean; for the beta family it is an approximation.
The simulation uses CL2 for the beta and negative-binomial responses.
Families that define their own sandwich (multinomial) have no
cluster-robust covariance.

\subsection{Asymptotic validity for fixed smoothing
parameters}\label{asymptotic-validity-for-fixed-smoothing-parameters}

\textbf{Proposition A.1.} \emph{Assume (i) the curves
\((\mathbf y_i, \mathbf X_i)\), \(i = 1,\dots,G\), are independent and
identically distributed with \(D_i \le \bar D < \infty\), with arbitrary
dependence within a curve; (ii) the basis, hence \(p\), is fixed and the
link is canonical; (iii) \(\boldsymbol\lambda = \boldsymbol\lambda_G\)
is non-random with \(\mathbf S_G/G \to \mathbf S_\infty\); (iv) the
weighted design rows are uniformly bounded, the curve scores have
bounded moments of order \(2 + \delta\) and dominated derivatives near
the target, and
\(\mathbf J = \mathrm E(\tilde{\mathbf X}_1^\top\tilde{\mathbf X}_1) + \mathbf S_\infty\)
is positive definite at
\(\boldsymbol\theta^*_\infty = \lim_G \boldsymbol\theta^*_G\), where
\(\boldsymbol\theta^*_G\) solves}
\begin{equation}\phantomsection\label{eq-s1-target}{
\mathrm E\,\mathbf u_1(\boldsymbol\theta) = (\mathbf S_G/G)\,\boldsymbol\theta .
}\end{equation} \emph{Then (a)
\(\sqrt G(\hat{\boldsymbol\theta} - \boldsymbol\theta^*_G) \to N(\mathbf 0, \mathbf J^{-1}\mathbf M\mathbf J^{-1})\)
with
\(\mathbf M = \mathrm{Var}\,\mathbf u_1(\boldsymbol\theta^*_\infty)\).
(b) With
\(\boldsymbol\mu = \mathrm E\,\mathbf u_1(\boldsymbol\theta^*_\infty) = \mathbf S_\infty\boldsymbol\theta^*_\infty\),
the variance part of Equation~\ref{eq-s1-cl2} (and of CR1) satisfies}
\begin{equation}\phantomsection\label{eq-s1-cov}{
G\,\frac{G}{G-1}\,\mathbf V_p\Bigl(\sum_i\mathbf U_i\mathbf U_i^\top\Bigr)\mathbf V_p \;\to_p\; \mathbf J^{-1}(\mathbf M + \boldsymbol\mu\boldsymbol\mu^\top)\mathbf J^{-1},
\qquad
G(\mathbf V_p - \mathbf V_e) \to \mathbf J^{-1}\mathbf S_\infty\mathbf J^{-1},
}\end{equation} \emph{so the variance part is consistent for the
covariance in (a) if
\(\mathbf S_\infty\boldsymbol\theta^*_\infty = \mathbf 0\) and
conservative otherwise; the Bayesian form adds
\(\mathbf J^{-1}\mathbf S_\infty\mathbf J^{-1}\). (c) If the mean model
is correct and
\(\sqrt G(\boldsymbol\theta^*_G - \boldsymbol\theta) \to \mathbf 0\)
(which forces \(\mathbf S_\infty\boldsymbol\theta = \mathbf 0\) even if
\(\mathbf S_\infty \neq \mathbf 0\)), the frequentist CL2 interval for
any fixed contrast is asymptotically valid for the true
\(\boldsymbol\theta\), and the Bayesian one is valid and conservative,
asymptotically exact if \(\mathbf S_\infty = \mathbf 0\), provided the
limiting variance
\(\mathbf a^\top\mathbf J^{-1}\mathbf M\mathbf J^{-1}\mathbf a\) of the
contrast is positive.}

The weights in \(\tilde{\mathbf X}_1\) are evaluated at
\(\boldsymbol\theta^*_\infty\), and a dispersion or other nuisance
parameter of the family is taken as known. The covariance in (a) is
unconditional: with random covariates, \(\mathbf M\) includes the
variation over curves of the conditional mean of the score, which is not
zero when \(\boldsymbol\theta^*_\infty\) differs from
\(\boldsymbol\theta\). Appendix A.2 instead conditions on the design,
where the same term appears as an upward bias of the meat.

For Gaussian responses with identity link, the finite-sample target
given the design is
\(\mathrm E(\hat{\boldsymbol\theta}) = (\mathbf I - \mathbf V_p\mathbf S)\boldsymbol\theta\),
of which Equation~\ref{eq-s1-target} is the population version; for GLMs
the target solves the expected penalized score equation.

\emph{Proof.} \(\hat{\boldsymbol\theta}\) solves an M-estimating
equation with independent curve contributions and a deterministic
penalty offset whose expectation vanishes at \(\boldsymbol\theta^*_G\);
(a) follows from the standard argument for such equations
(\citeproc{ref-LiangZeger1986}{Liang and Zeger 1986};
\citeproc{ref-HansenLee2019}{Hansen and Lee 2019}). For (b), bounded
curve sizes and fixed \(p\) make the leverages \(O_p(G^{-1})\), so CL2
and CR1 meats agree to first order, and the meat converges to the second
moment of a score with mean \(\boldsymbol\mu\). For (c), a correct mean
model gives curve scores with mean zero whatever the dependence within
the curve. \(\square\)

The proposition does not cover smoothing parameters selected from the
data, nor a number of coefficients comparable to or larger than \(G\);
both hold in our applications and simulations.

\subsection{Covariance components}\label{covariance-components}

\begin{figure}

\centering{

\pandocbounded{\includegraphics[keepaspectratio]{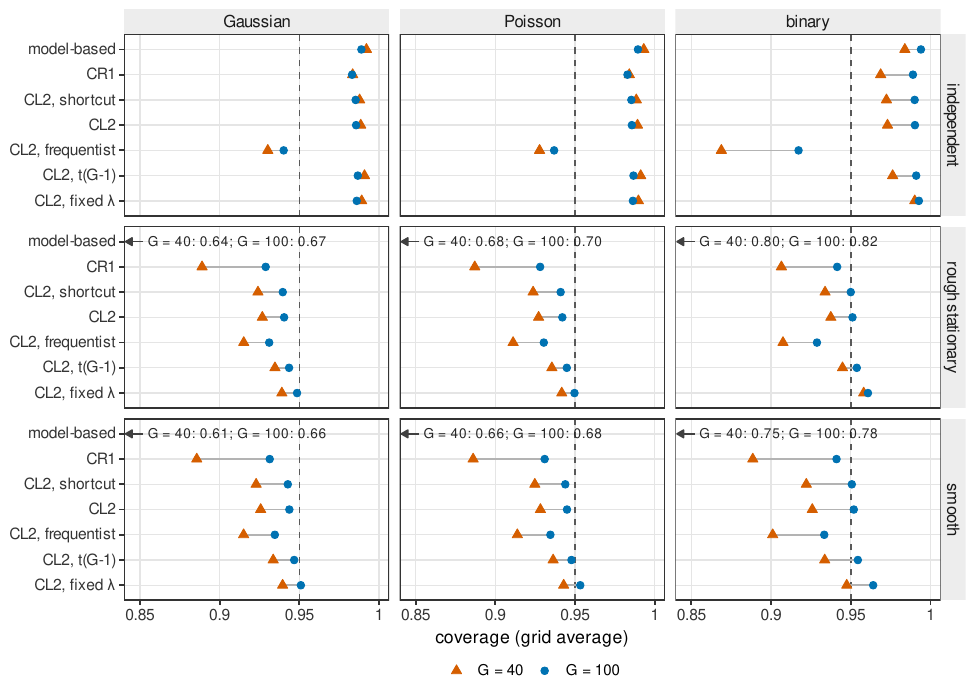}}

}

\caption{\label{fig-s1-beta}Covariance components, \(\beta(s,t)\):
grid-averaged coverage of nominal 95\% pointwise intervals by family,
error process and number of curves \(G\), main factorial settings at the
default signal level, 200 replicates. Model-based: \(\mathbf V_c\); CR1,
CL2, shortcut (CL2 with the \(\mathbf I - \tilde{\mathbf H}_{ii}\)
approximation) and CL2 in the Bayesian form; CL2, frequentist:
frequentist form; CL2, t(G-1): CL2 with \(t\) critical value; CL2, fixed
\(\lambda\): CL2 after a refit with smoothing parameters fixed at the
setting's geometric-mean REML values. Arrows at the left edge mark
coverage below the axis range; the values are printed next to them.
Dashed: 0.95.}

\end{figure}

\begin{figure}

\centering{

\pandocbounded{\includegraphics[keepaspectratio]{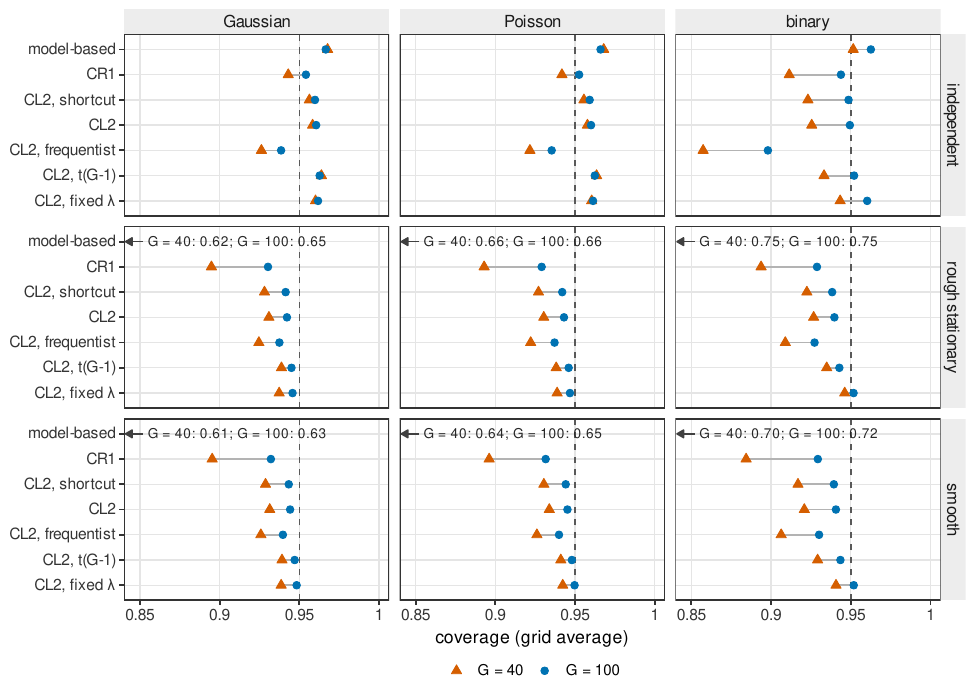}}

}

\caption{\label{fig-s1-mean}Covariance components, mean
\(\mathrm E(Y \mid X)\): grid-averaged coverage of nominal 95\%
pointwise intervals; intervals, settings and marks as in
Figure~\ref{fig-s1-beta}.}

\end{figure}

\begin{figure}

\centering{

\pandocbounded{\includegraphics[keepaspectratio]{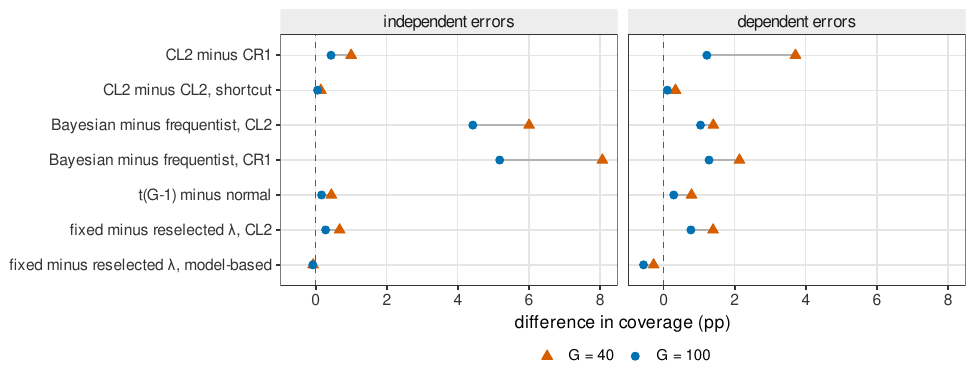}}

}

\caption{\label{fig-s1-contrasts}Covariance components: paired coverage
differences in percentage points, averaged over families and over the
mean and \(\beta(s,t)\), by error process and number of curves \(G\).
Each difference compares two intervals on the same replicates; unless
stated, both are CL2 in the Bayesian form with the normal critical
value. Shortcut: the \(\mathbf I - \tilde{\mathbf H}_{ii}\)
approximation. The SE of the averages cannot be recovered from the
retained per-setting summaries, because the settings share replicates;
the range of the per-setting paired Monte Carlo SEs is in Supplement
S7.}

\end{figure}

\begin{figure}

\centering{

\pandocbounded{\includegraphics[keepaspectratio]{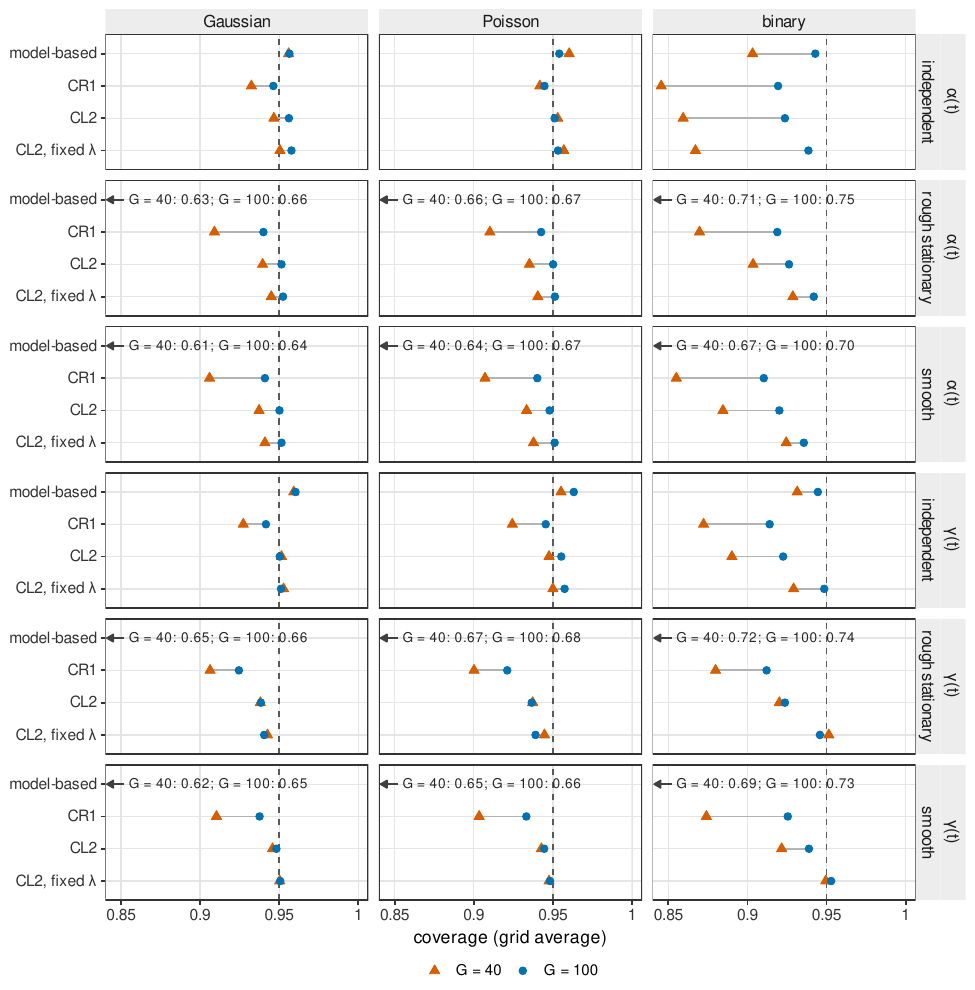}}

}

\caption{\label{fig-s1-alpha-gamma}Covariance components, intercept
\(\alpha(t)\) and coefficient function \(\gamma(t)\): grid-averaged
coverage of nominal 95\% pointwise intervals, settings and marks as in
Figure~\ref{fig-s1-beta}; model-based (\(\mathbf V_c\)), CR1 and CL2
(Bayesian form), and CL2 after a refit with smoothing parameters fixed
at the setting's geometric-mean REML values.}

\end{figure}

The comparison uses the 18 main factorial settings at the default signal
level (\(D =\) 61, \(G\) = 40 or 100, 200 replicates per setting) with
common random numbers. Except for the fixed-\(\lambda\) interval, every
covariance is computed for the same REML fit of each replicate; these
fits reproduce the stored fits of the main study to numerical precision.
The intervals are the model-based interval (\(\mathbf V_c\)); CR1, CL2
with the \(\mathbf I - \tilde{\mathbf H}_{ii}\) approximation, and CL2,
all in the Bayesian form of Equation~\ref{eq-s1-cl2}; CL2 in the
frequentist form; CL2 with the critical value \(t_{G-1,0.975}\) in place
of \(z_{0.975}\); and CL2 after a refit of the same data with the
smoothing parameters fixed, per penalty, at the geometric mean of the
setting's REML estimates. The fixed-\(\lambda\) refit removes the
replicate-to-replicate variability of the smoothing parameters and also
moves them to their typical level. Figure~\ref{fig-s1-beta},
Figure~\ref{fig-s1-mean} and Figure~\ref{fig-s1-alpha-gamma} show the
coverage, Figure~\ref{fig-s1-contrasts} the paired differences;
Supplement S7 lists the values.

Under dependence, the leverage adjustment adds 3.7 pp of coverage over
CR1 at \(G = 40\) and 1.2 pp at \(G = 100\); the full block adds at most
0.4 pp in any setting over the \(\mathbf I - \tilde{\mathbf H}_{ii}\)
approximation (under dependence 0.1--0.4 pp), in line with
\(\mathbf A^{\mathrm s}_i\mathbf R_i\mathbf A^{\mathrm s}_i \preceq \mathbf I\).
The allowance \(\mathbf V_p - \mathbf V_e\) matters under independent
errors (6.0 pp at \(G = 40\)), much less under dependence (1.4 pp).

With reselected smoothing parameters, the grid average of CL2's
root-mean-square standard error divided by the empirical standard
deviation of \(\hat\beta\) under dependence is 0.92 at \(G = 40\) and
0.97 at \(G = 100\); with fixed smoothing parameters it is 1.05 and
1.06, and CL2 coverage of \(\beta(s,t)\) rises from 0.929 to 0.945 at
\(G = 40\). Fixing the smoothing parameters, which
Equation~\ref{eq-s1-cl2} conditions on, at their typical level closes
most of the remaining shortfall at \(G = 40\); the \(t_{G-1}\) critical
value recovers 0.8 pp of it.

Fixed smoothing parameters also raise CL2 coverage of \(\alpha(t)\) and
\(\gamma(t)\) under dependence (Figure~\ref{fig-s1-alpha-gamma}): for
binary responses with smooth errors at \(G = 40\), CL2 coverage of
\(\alpha(t)\) rises from 0.885 to 0.925 and that of \(\gamma(t)\) from
0.922 to 0.950. Under independent errors, the binary \(\alpha(t)\) at
\(G = 40\) stays below nominal with fixed smoothing parameters (0.867).

\subsection{Satterthwaite degrees of
freedom}\label{satterthwaite-degrees-of-freedom}

For a contrast \(\mathbf a^\top\boldsymbol\theta\) the sandwich part of
Equation~\ref{eq-s1-cl2} is a quadratic form in the whitened residuals,
\(\mathbf a^\top\mathbf V_p(\sum_i\mathbf U_i\mathbf U_i^\top)\mathbf V_p\mathbf a = \sum_i (\mathbf q_i^\top\mathbf z_i)^2\)
with
\(\mathbf q_i = \mathbf R_i^{-1/2}\tilde{\mathbf X}_i\mathbf V_p\mathbf a\)
(times \(G/(G-1)\)). Under the Gaussian working model with fixed
smoothing parameters and no smoothing bias
(\(\mathbf S\boldsymbol\theta = \mathbf 0\), so that the residual mean
\(\mathbf m\) of Equation~\ref{eq-s1-mean} vanishes),
\(\mathbf z = (\mathbf I - \tilde{\mathbf H})\boldsymbol\varepsilon\)
with independent errors. The quadratic form then has mean
\(\operatorname{tr}\boldsymbol\Gamma\) and variance
\(2\operatorname{tr}(\boldsymbol\Gamma^2)\), where the \(G \times G\)
matrix \(\boldsymbol\Gamma\) has entries
\(\Gamma_{ih} = \mathbf q_i^\top[(\mathbf I - \tilde{\mathbf H})(\mathbf I - \tilde{\mathbf H})^\top]_{ih}\mathbf q_h\).
Matching a scaled \(\chi^2_\nu\) gives the Satterthwaite degrees of
freedom \begin{equation}\phantomsection\label{eq-s1-satt}{
\nu(\mathbf a) = \frac{(\operatorname{tr}\boldsymbol\Gamma)^2}{\operatorname{tr}(\boldsymbol\Gamma^2)}
}\end{equation} (\citeproc{ref-BellMcCaffrey2002}{Bell and McCaffrey
2002}; \citeproc{ref-PustejovskyTipton2018}{Pustejovsky and Tipton
2018}), computed for every evaluation point; the interval uses
\(t_{\nu, 0.975}\) in place of \(z_{0.975}\). The Satterthwaite
intervals evaluated here compute Equation~\ref{eq-s1-satt} for the
sandwich part only. A variant matches the moments of the full Bayesian
CL2 variance, whose term
\(\mathbf a^\top(\mathbf V_p - \mathbf V_e)\mathbf a\) is fixed, which
increases \(\nu\). Using only the diagonal of \(\boldsymbol\Gamma\)
gives practically the same \(\nu\).

With the floor inactive,
\(\Gamma_{ii} = \mathbf q_i^\top\mathbf R_i\mathbf q_i = \lVert\tilde{\mathbf X}_i\mathbf V_p\mathbf a\rVert^2\)
and
\(\operatorname{tr}\boldsymbol\Gamma = \mathbf a^\top\mathbf V_e\mathbf a\):
\(\Gamma_{ii}\) is the contribution of curve \(i\) to the working-model
variance of the contrast, and the diagonal version of \(\nu\), an upper
bound, is \((\sum_i\Gamma_{ii})^2/\sum_i\Gamma_{ii}^2\), an effective
number of curves behind the contrast (compare
\citeproc{ref-ImbensKolesar2016}{Imbens and Kolesár 2016}). For the
intercept, the curves contribute more equally and \(\nu\) is larger. For
a contrast that weights curves by a single covariate value or score
\(x_i\), as for \(\gamma(t)\), \(\Gamma_{ii}\) is roughly proportional
to \(x_i^2\), and for Gaussian \(x_i\),
\((\sum_i x_i^2)^2/\sum_i x_i^4 \approx G/3\). Contrasts that involve
several scores, as for \(\beta(s,t)\) and fitted means, tend to give
somewhat more. The degrees of freedom are smaller with heavy-tailed
covariates or single curves of high leverage, which the covariates of
our simulation do not include.

The Satterthwaite intervals use the REML fits and CL2 covariances of
Appendix A.6 (the 18 main factorial settings at the default signal
level, 200 replicates); Section~\ref{sec-res-intervals},
Section~\ref{sec-plasmode-results} and Supplement S1.2, S1.7 and S2.1
give them with 20 curves, in the extensions and on the plasmode data.

The median per-point degrees of freedom are 12--30 with \(G = 40\)
curves and 30--87 with \(G = 100\); across both, that is 0.30--0.88
times \(G - 1\).

\begin{figure}

\centering{

\pandocbounded{\includegraphics[keepaspectratio]{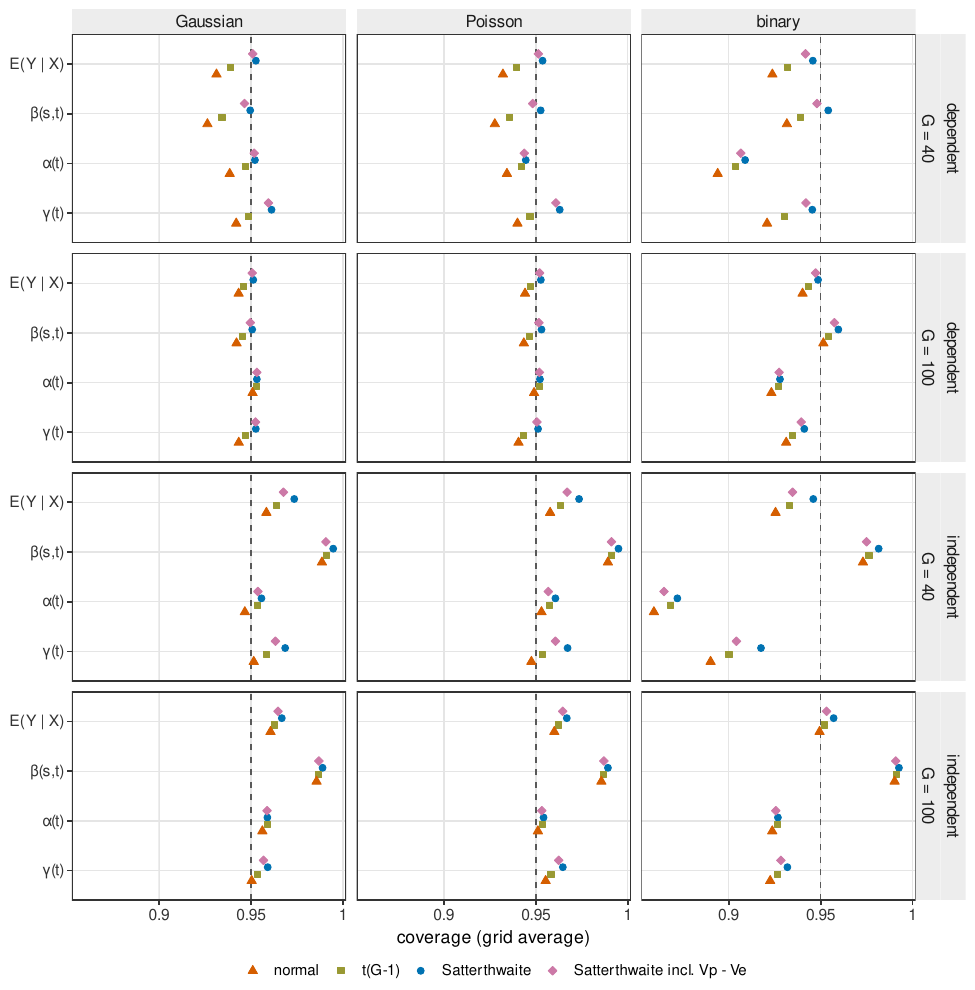}}

}

\caption{\label{fig-s1-satt}Critical values for REML + CL2:
grid-averaged coverage by dependence, number of curves \(G\), family and
estimand; means over the error processes of the main factorial at the
default signal level, 200 replicates. Satterthwaite: per-point degrees
of freedom for the sandwich part; incl.~Vp - Ve: degrees of freedom for
the full Bayesian CL2 variance. Dashed: 0.95. The 5\% quantile of
pointwise coverage and the width relative to the normal critical value
are in Supplement S7.}

\end{figure}

Under dependent errors, Satterthwaite intervals cover the mean,
\(\beta(s,t)\) and \(\gamma(t)\) at 0.95--0.96 with 40 curves (normal
critical value: 0.92--0.94) and at 0.94--0.96 with 100 (0.93--0.95), at
1.10--1.12 and 1.04--1.05 times the width with the normal critical value
(Figure~\ref{fig-s1-satt}).

Satterthwaite intervals do not repair the binary intercept: they cover
0.91--0.93 under dependence and 0.87--0.93 under independent errors. For
this term, the Satterthwaite degrees of freedom are large (the intervals
widen by a factor of only 1.01--1.05), so the reference distribution
cannot make up for a standard error that is too small under independent
errors or for the additional shift of the studentized error under
dependence (Section~\ref{sec-res-intervals}, Supplement S1.4).

Under independent errors, Satterthwaite intervals overcover \(\beta\)
further (0.98--0.99). The variant whose degrees of freedom include
\(\mathbf V_p - \mathbf V_e\) lies between \(t_{G-1}\) and the plain
Satterthwaite reference.

\emph{Non-Gaussian responses.} We use the same \(\nu\) for all families
(Section~\ref{sec-cl2}), although its derivation assumes Gaussian
errors. Let the entries of \(\boldsymbol\varepsilon\) be independent
standardized score contributions with excess kurtosis \(\kappa_k\) (for
Poisson responses, \(\kappa_k = 1/\mu_k\)), and let \(\mathbf Q\) be the
matrix with \(G\) columns whose \(i\)th column holds \(\mathbf q_i\) in
the rows of curve \(i\) and zeros elsewhere. With
\(\mathbf B = (\mathbf I - \tilde{\mathbf H})^\top\mathbf Q\mathbf Q^\top(\mathbf I - \tilde{\mathbf H})\),
the quadratic form has, to first order, the same mean
\(\operatorname{tr}\boldsymbol\Gamma\) but variance
\(2\operatorname{tr}(\boldsymbol\Gamma^2) + \sum_k \kappa_k B_{kk}^2\).
Positive excess kurtosis therefore makes Equation~\ref{eq-s1-satt} too
large and the interval too narrow, and negative excess kurtosis does the
reverse. The size of the extra term depends on
\(\sum_k B_{kk}^2/\operatorname{tr}(\boldsymbol\Gamma^2) \le 1\). If
\(\tilde{\mathbf H}\) is ignored, this ratio is \(1/D\) when all entries
of the \(\mathbf q_i\) have the same magnitude, so we expect little
effect for contrasts that load on many grid points, unless \(\kappa_k\)
is comparable to \(D\). Sparse grids, localized contrasts and rare
binary events are the less favorable cases. We did not compute this term
or evaluate a family-dependent \(\nu\). Empirically, under dependent
errors with 40 curves, Satterthwaite intervals cover the mean, \(\beta\)
and \(\gamma\) at 0.95--0.96 for Gaussian, 0.95--0.96 for Poisson and
0.95 for binary responses.

\section{Setups of the two studies}\label{sec-app-setup}

\subsection{Simulation design}\label{simulation-design}

The data follow the model in Equation~\ref{eq-model} with
\(i = 1, \dots, G\) curves. Responses are generated on a grid of 241
points in \(t \in [0, 1]\) and subsampled to \(D = 61\) (default) or 241
points; the functional covariate is observed on 51 points in
\(s \in [0, 1]\):
\(X_i(s) = \xi_{i0} + \sum_{k=1}^{29} \xi_{ik}\sqrt 2 \cos(k\pi s)\)
with independent \(\xi_{ik} \sim N(0, (k+1)^{-2})\). Its centered
spectrum (first principal component 62\% of the variance, four
components 88\%, rank 24 at 99.5\%) matches the corpus-callosum profiles
of the baseline DTI data in refund (67\%, 86\%, 26). The low-rank
variant uses variances \((k+1)^{-3}\) (rank 9). The scalar covariate is
\(v_i \sim N(0, 1)\). The smooth-effect model replaces the functional
term by \(f(x_i, t)\) with \(x_i \sim U(0, 1)\).

Truths are fixed functions projected by least squares onto the fitted
bases, so they are exactly representable:
\(\beta(s,t) = \cos(\pi s)\sin(\pi t) + (s - 1/2)\cos(2\pi t)\),
\(f(x,t) = 2(x - 1/2)\cos(\pi t) + t\sin(\pi x)\),
\(\gamma(t) = \sin(2\pi t)\). The rough truth adds
\(0.6\sin(3\pi s)\sin(3\pi t)\) to \(\beta\) and \(f\) and
\(0.5\sin(5\pi t)\) to \(\gamma\). Truths are scaled so that the
non-intercept signal
\(S = \int \mathrm{Var}_{X,v}\{\eta(t) - \alpha(t)\}\,dt\) is 1 (80\%
from the main term, 20\% from \(\gamma\)) and multiplied by \(\sqrt S\)
per family. The intercept is a Gaussian bump plus a constant that sets
the average response level (mean count 3, prevalence 0.2).

The fitted bases are cubic B-splines with first-order difference
penalties (\citeproc{ref-Eilers1996}{Eilers and Marx 1996};
\citeproc{ref-Wood2017}{Wood 2017}) (mgcv's \texttt{ps} basis with
\texttt{m\ =\ c(2,\ 1)}), so constants along each direction are
unpenalized; \(k = 12\) along \(t\) for the intercept and \(\gamma\); an
\(8 \times 10\) tensor product for \(\beta\); \(8 \times 12\) for \(f\).
The larger bases halve and quarter the knot spacing (about 100, 260 and
790 coefficients).

\begin{figure}

\centering{

\pandocbounded{\includegraphics[keepaspectratio]{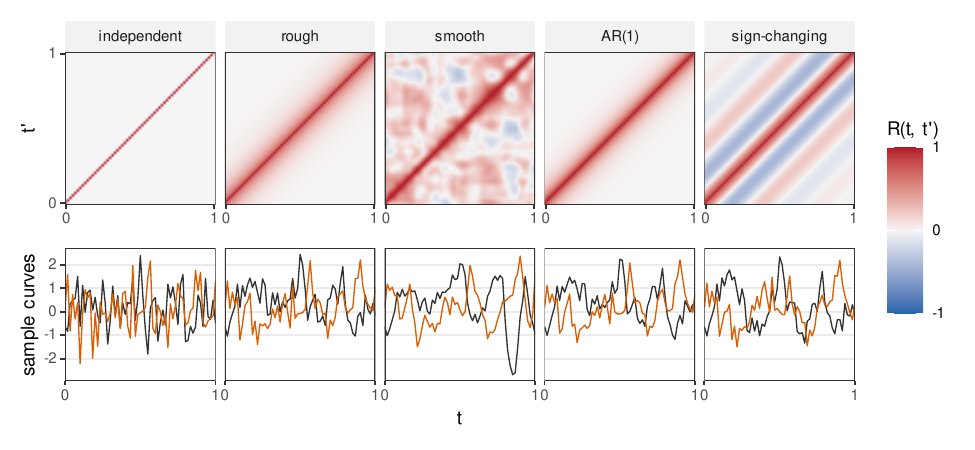}}

}

\caption{\label{fig-kernels}Error processes of the simulation:
correlation kernels \(R(t,t')\) (top) and two sample curves of the
latent Gaussian process (bottom, same draws of standard normals across
processes), on the default grid of \(D = 61\) points. The main factorial
uses the first three.}

\end{figure}

Within-curve errors are a zero-mean process with correlation kernel
\(R\), scaled for Gaussian responses and used as the latent layer of a
Gaussian copula for the other families,
\(Y_{ij} = F^{-1}_{ij}\{\Phi(Z_{ij})\}\), so each marginal is exactly
the fitted family and the working model is wrong only about the
dependence. The kernels are: independent; rough stationary
\(0.8\exp(-|t-t'|/0.089) + 0.2\,\mathbb 1\{t=t'\}\); smooth: the
residual covariance of a REML fit to the DTI data (24 eigenfunctions
plus 6\% white noise, as a correlation kernel); exact AR(1)
\(\exp(-|t-t'|/0.071)\); sign-changing
\(0.8\exp(-|t-t'|/0.3)\cos(2\pi|t-t'|/0.4) + 0.2\,\mathbb 1\{t=t'\}\);
heteroskedastic versions of the smooth kernel with the DTI variance
profile along \(t\), a subject scale \(\exp(0.5 v_i - 0.25)\), or both.
Figure~\ref{fig-kernels} shows the homoskedastic kernels.

Under misregistration,
\(Y_i(t_j) = \eta_i(h_i(t_j)) + \varepsilon_i(h_i(t_j))\) with
\(h_i(t) = t + a_i\sin(\pi t)/\pi\), \(a_i \sim N(0, 0.314^2)\)
truncated at \(|a_i| \le 0.9\), and errors are independent. Gaussian
misregistration settings score the warp-averaged coefficient functions,
GLM settings only the exact observed-clock mean.

The signal is \(R^2 \in \{0.2, 0.5, 0.8\}\) for Gaussian responses and
\(\sqrt S \in \{0.5, 1\}\) on the link scale for Poisson and binary
responses. The extended families are run at the default signal: scaled
\(t\) (4 df) as Gaussian at \(R^2 = 0.5\); beta (\(\phi = 20\), mean
0.3) and negative binomial (size 5, mean 3) at \(\sqrt S = 0.5\).

The 20-curve block uses the first 20 curves of each replicate of the
main factorial, so it shares its draws with the 40- and 100-curve
settings. The zero-surface block sets \(\beta \equiv 0\) in the
smooth-truth setting at the default signal, for Gaussian and binary
responses with independent and smooth errors and 40 and 100 curves;
\(\gamma\), the intercept bump and the Gaussian error variance are those
of the non-zero setting, and the intercept constant of binary responses
is reset to keep the prevalence at 0.2.

All random draws depend only on the replicate index, so contrasts
between settings and intervals are paired. Each setting has 200
replicates; none failed. Concurvity of the fitted terms (mgcv's worst
case) is at most 0.25. The next table lists the blocks of settings: the
main factorial (family × error process × \(G\) × signal) and extensions
that each change one factor.

\begin{landscape}\begin{table}[H]
\centering
\caption{\label{tab:s2-design}Blocks of the simulation design. Error processes: indep. independent, rough rough stationary, smooth from the DTI residuals, AR(1) exact AR(1), sign-chg. sign-changing, var(t)/var(z)/var(t,z) heteroskedastic versions of smooth, misreg. misregistration. Truth: smooth, rough or zero. Signal: low, default, high. Bases: default, 2× or 4× finer knots.}
\centering
\resizebox{\ifdim\width>\linewidth\linewidth\else\width\fi}{!}{
\fontsize{7}{9}\selectfont
\begin{tabular}[t]{lrlllllllll}
\toprule
block & settings & families & errors & G & signal & truth & D & model & covariate & basis\\
\midrule
main factorial & 42 & Gaussian, Poisson, binary & indep., rough, smooth & 40, 100 & low, default, high & smooth & 61 & functional & rich & default\\
main factorial, G = 20 & 21 & Gaussian, Poisson, binary & indep., rough, smooth & 20 & low, default, high & smooth & 61 & functional & rich & default\\
misregistration & 12 & Gaussian, Poisson, binary & misreg. & 40, 100 & high & smooth, rough & 61 & functional & rich & default\\
dense grid & 9 & Gaussian, Poisson, binary & indep., rough, smooth & 100 & default & smooth & 241 & functional & rich & default\\
rough truth & 14 & Gaussian, Poisson, binary & indep., smooth & 100 & low, default, high & rough & 61 & functional & rich & default\\
zero surface & 8 & Gaussian, binary & indep., smooth & 40, 100 & default & zero & 61 & functional & rich & default\\
smooth effect & 9 & Gaussian, Poisson, binary & indep., rough, smooth & 100 & default & smooth & 61 & smooth effect & rich & default\\
extended families & 8 & scaled t, beta, negative binomial, negative binomial (fitted as Poisson) & indep., smooth & 100 & default & smooth & 61 & functional & rich & default\\
exact AR(1) & 1 & Gaussian & AR(1) & 100 & default & smooth & 61 & functional & rich & default\\
sign-changing & 3 & Gaussian, Poisson, binary & sign-chg. & 100 & default & smooth & 61 & functional & rich & default\\
misreg. (AR(1) fit) & 1 & Gaussian & misreg. & 100 & high & rough & 61 & functional & rich & default\\
heteroskedastic & 6 & Gaussian & var(t), var(z), var(t,z) & 40, 100 & default & smooth & 61 & functional & rich & default\\
low-rank covariate & 6 & Gaussian, Poisson, binary & indep., smooth & 100 & default & smooth & 61 & functional & low-rank & default\\
larger bases & 12 & Gaussian, Poisson, binary & indep., smooth & 100 & default & smooth & 61 & functional & rich & 2× finer knots, 4× finer knots\\
\bottomrule
\end{tabular}}
\end{table}
\end{landscape}

Each replicate is fitted by REML and curve-blocked NCV (mgcv
\texttt{gam}); by pointwise (non-blocked) NCV in the Gaussian settings
at \(G = 100\) with independent, smooth and sign-changing errors; and by
the AR(1) working model in the Gaussian settings at \(G = 100\) with
dependent errors. The AR(1) model is fitted with mgcv's \texttt{bam()}
by fast REML with AR(1) errors that restart at the first grid point of
each curve. Its correlation \(\rho\) is profiled over the grid
\(\{0, 0.1, \dots, 0.9, 0.95, 0.99\}\) by the fast-REML criterion, which
includes the log-determinant of the whitening transform; the best grid
value is then refined by a one-dimensional search between its two
neighboring grid values. Its intervals are model-based from that fit.
The plasmode study uses the same fit.

Some NCV fits were flagged as not converged, meaning that mgcv's search
over the smoothing parameters ended without meeting its convergence
test: 88\% of scaled-\(t\) fits under smooth errors and 68\% under
independent errors, 1.0\% of beta fits, and fewer than 0.1\% otherwise
(settings with at least 40 curves and a non-zero surface; at most 2.5\%
per combination in the plasmode study). No REML fit in these settings
was flagged, so the study says nothing about intervals from
non-converged REML fits.

We refitted flagged scaled-\(t\) fits from the REML smoothing parameters
and from ten times those values. For the first five flagged fits, all
with independent errors, the refits reach the same criterion within
\(8 \times 10^{-5}\) and the same surface within \(6 \times 10^{-4}\).
Of 20 randomly chosen flagged fits with smooth errors, 17 reach the same
criterion within 0.005, the same surface within 0.004 and the same CL2
standard errors within 0.3\%, consistent with a flat optimum; in an 18th
the study fit's criterion is 2.2 above the refits', with the surface
within 0.04. In the other two, a refit returns a materially different
fit, its surface moved by about twice the root-mean-square size of the
true surface (1.8). In one, the study fit has the worse criterion, by
17, and the standard errors differ by up to 80\%; in the other, a refit
is worse by 10, with standard errors up to 403\% larger. 38 of the 40
refits were flagged too. A flag thus does not separate these fits from
the stable ones.

Under smooth errors, in the 176 replicates with a flagged NCV fit, the
NCV estimate of \(\beta\) has median relative error 0.13 and NCV + CL2
covers 0.938 (Monte Carlo SE 0.004), against 0.12 and 0.959 (0.008) in
the 24 other replicates; this difference of 2.1 pp is the largest
coverage gap of NCV + CL2 over estimands and error processes. The
model-based NCV intervals cover 0.697 and 0.742, REML + CL2 0.943 and
0.951 in the same two groups, so part of the gap may reflect harder
data. The flag is not randomly assigned and the comparison is
descriptive.

Low-smoothing NCV fits (Section~\ref{sec-res-estimation}) were examined
in the main factorial and in the settings with denser grids, other
families, rough truths, AR(1) errors and misregistration (86 settings
with 40 or 100 curves), with the tail defined on the first surface
penalty. Under dependence, apart from misregistration, cutoffs of
\(e^4\) and \(e^2\) give 4\% and 9\% of NCV fits, and applying the
\(e^3 \approx 20\) cutoff to either penalty gives 7.4\%. Under
misregistration, NCV's smoothing parameters are the more variable ones,
with up to 24\% of fits below the \(e^3\) cutoff (Poisson). Under
dependence, apart from misregistration, the share of low-smoothing fits
is 4.0--7.9\% across families and error processes, and these fits have
1.4 times their setting's median surface EDF, still fewer than REML's on
all of them. REML also does worse on these replicates: its error is 1.6
times its median and its CL2 intervals cover 0.858 (NCV + CL2: 0.831).
On the remaining replicates, REML's mean squared error is 8.9 rather
than 6.9 times NCV's (geometric means over settings). The converse does
not hold: on the 17\% of replicates in REML's own lower tail, NCV's
error stays near its median (1.05 times) and its coverage is 0.6 pp
lower than on the other replicates.

\subsection{Plasmode datasets}\label{plasmode-datasets}

A plasmode simulation
(\citeproc{ref-FranklinSchneeweissPolinskiRassen2014}{Franklin et al.
2014}) keeps the real covariates and real error structure of a dataset
and replaces only the true regression function by a known one.

\textbf{ECG strain} (independent curves; myocardial-strain study, first
beat per subject; non-public; aggregates only). 78 curves; response:
global longitudinal strain, 4-chamber view (\%), one cardiac cycle
normalized to {[}0, 1{]}, native grid about 52 frames per beat (29--83)
(fitted: interpolated to 61 points); functional covariate:
simultaneously recorded ECG of the same beat (digitized pixel intensity,
shape usable, amplitude up to scale), 51 points; scalar covariate: age
(standardized). Preprocessing: cycle time normalized to the R--R
interval; subjects with at least two beats (the second beat supplies the
beat-to-beat difference curves). REML fit: \(R^2\) 0.47, surface EDF
59.6 (NCV fit: 44.0); residual design effect 19.9, lag-one correlation
0.96, 15\% negative correlations, 4 principal components for 90\% of the
residual variance, covariate rank 15.

\textbf{AF trial} (independent curves; atrial-fibrillation trial,
patients randomized to digoxin or a beta-blocker; non-public; aggregates
only). 118 curves; response: left-atrial strain, 4-chamber view, at day
30 (\%), one cycle on {[}0, 1{]}, native grid 30 points per cycle
(fitted: used as is); functional covariate: the same strain curve at day
0, 51 points; scalar covariate: randomized treatment (digoxin = 1).
Preprocessing: none beyond the cycle normalization; a baseline-adjusted
functional ANCOVA. REML fit: \(R^2\) 0.19, surface EDF 33.2 (NCV fit:
25.3); residual design effect 10.9, lag-one correlation 0.89, 20\%
negative correlations, 3 principal components for 90\% of the residual
variance, covariate rank 8.

\textbf{running} (independent curves; Fukuchi, Fukuchi \& Duarte 2017
(PeerJ 5:e3298); one row per person, first condition; public
(tidyfundata, MIT; original data public)). 90 curves; response: sagittal
knee moment over normalized stance, native grid 101 points (fitted:
interpolated to 61 points); functional covariate: sagittal knee angle
over the same stance, 51 points; scalar covariate: body mass
(standardized). Preprocessing: one row per person (post-intervention
rows dropped); speed and backpack conditions differ between persons and
stay in the residuals. REML fit: \(R^2\) 0.46, surface EDF 53.7 (NCV
fit: 54.8); residual design effect 33.3, lag-one correlation 0.99, 12\%
negative correlations, 2 principal components for 90\% of the residual
variance, covariate rank 7.

\textbf{DTI} (independent curves; refund::DTI, baseline visit (Goldsmith
et al.~2011): multiple-sclerosis patients and controls; public (refund,
GPL ≥ 2)). 92 curves; response: fractional anisotropy along the right
corticospinal tract, native grid 55 points (fitted: used as is);
functional covariate: fractional anisotropy along the corpus callosum,
51 points; scalar covariate: case status (MS = 1). Preprocessing:
complete cases of both tracts. REML fit: \(R^2\) 0.21, surface EDF 53.8
(NCV fit: 35.1); residual design effect 9.9, lag-one correlation 0.81,
19\% negative correlations, 16 principal components for 90\% of the
residual variance, covariate rank 24.

\textbf{gait} (independent curves; Van Criekinge et al.~2023 (Sci Data
10:852; figshare 24192489), able-bodied adults walking barefoot at
preferred speed; public (CC0)). 138 curves; response: sagittal knee
moment over one stride, one mean stride per subject, native grid 1001
points (0.1\% of the stride) (fitted: every 10th point (101));
functional covariate: sagittal knee angle over the stride, 51 points;
scalar covariate: age (standardized; from the subject-characteristics
file). Preprocessing: the authors' stride normalization and averaging
over strides (within-subject stride variability is averaged out). REML
fit: \(R^2\) 0.40, surface EDF 68.7 (NCV fit: 60.8); residual design
effect 37.0, lag-one correlation 0.98, 33\% negative correlations, 6
principal components for 90\% of the residual variance, covariate rank
8.

\textbf{ECG 8-lead} (independent curves; roahd::mfD\_healthy and
mfD\_LBBB (Ieva et al.), 50 healthy and 50 left-bundle-branch-block
subjects; public (roahd, GPL-3)). 100 curves; response: ECG lead V5 (µV)
over one beat, native grid 1024 points at 1 kHz (fitted: interpolated to
101 points); functional covariate: ECG lead I, 51 points; scalar
covariate: LBBB indicator. Preprocessing: registered and smoothed by the
package authors. REML fit: \(R^2\) 0.21, surface EDF 39.4 (NCV fit:
11.2); residual design effect 6.2, lag-one correlation 0.94, 32\%
negative correlations, 4 principal components for 90\% of the residual
variance, covariate rank 15.

\textbf{ocean} (dependent curves; Hawaii Ocean Time-series profiles
(FRegSigCom::ocean), monthly casts; public (FRegSigCom, GPL ≥ 2; HOT
data public)). 116 curves; response: dissolved oxygen over depth, native
grid 101 points (fitted: used as is); functional covariate: temperature
over depth, 51 points; scalar covariate: cast order (standardized).
Preprocessing: none; casts kept in time order. REML fit: \(R^2\) 0.36,
surface EDF 59.2 (NCV fit: 49.8); residual design effect 39.7, lag-one
correlation 0.99, 22\% negative correlations, 5 principal components for
90\% of the residual variance, covariate rank 6.

\textbf{weather} (dependent curves; AEMET open data via fda.usc::aemet,
73 Spanish stations, daily means 1980--2009; public (fda.usc, GPL-2)).
73 curves; response: log precipitation over the year, native grid 365
days (fitted: every 5th day (73)); functional covariate: daily mean
temperature over the year, 51 points; scalar covariate: altitude
(standardized). Preprocessing: none beyond the day thinning. REML fit:
\(R^2\) 0.41, surface EDF 29.5 (NCV fit: 38.6); residual design effect
29.1, lag-one correlation 0.50, 1\% negative correlations, 19 principal
components for 90\% of the residual variance, covariate rank 4.

\textbf{electricity} (dependent curves; Adelaide electricity demand
(fds::mondaydemand, fds::mondaytempairport; Magnano, Boland \& Hyndman
2008), every 5th Monday 1997--2007; public (fds, GPL-3)). 102 curves;
response: half-hourly electricity demand over the day, native grid 48
half-hours (fitted: used as is); functional covariate: airport
temperature over the same day, 51 points; scalar covariate: week index
(standardized; absorbs the linear trend). Preprocessing: every 5th
Monday (thins, not removes, the week-to-week dependence; the annual
cycle stays). REML fit: \(R^2\) 0.31, surface EDF 33.2 (NCV fit: 10.0);
residual design effect 35.8, lag-one correlation 0.98, 0\% negative
correlations, 2 principal components for 90\% of the residual variance,
covariate rank 11.

The ECG strain beats were recorded with 29 to 83 frames and interpolated
to 61 grid points; interpolation adds smoothness within a curve, so part
of the residual dependence of that dataset, and of its design effect,
comes from the interpolation.

Six datasets have independent source curves. The other three have
serially (ocean, electricity) or spatially (weather) dependent source
curves.

The residual \emph{design effect} of a dataset is
\(\mathbf 1^\top\boldsymbol\Sigma\mathbf 1/\operatorname{tr}\boldsymbol\Sigma\)
for the covariance \(\boldsymbol\Sigma\) of its REML residual curves:
the factor by which the variance of a grid mean exceeds its
working-independence value.

\begin{table}[H]
\centering
\caption{\label{tab:t-s3-models-1}Fitted models: bases (intercept, $\gamma$; surface $8 \times 10$), $R^2$ of the REML fit, surface and total EDF of the three fits used as truths (REML / NCV / midpoint), and whether the midpoint fit lies between the two for every term.}
\centering
\resizebox{\ifdim\width>\linewidth\linewidth\else\width\fi}{!}{
\fontsize{6}{8}\selectfont
\begin{tabular}[t]{lrrlrlll}
\toprule
dataset & intercept basis k & gamma(t) basis k & surface basis (s × t) & R² & surface EDF REML / NCV / midpoint & total EDF REML / NCV / midpoint & midpoint interior\\
\midrule
ECG strain & 12 & 12 & 8 × 10 & 0.469 & 59.6 / 44.0 / 52.9 & 77.1 / 54.1 / 63.4 & FALSE\\
AF trial & 12 & 12 & 8 × 10 & 0.192 & 33.2 / 25.3 / 29.5 & 44.7 / 32.8 / 38.9 & TRUE\\
running & 12 & 12 & 8 × 10 & 0.458 & 53.7 / 54.8 / 60.4 & 68.0 / 65.9 / 73.1 & FALSE\\
DTI & 12 & 12 & 8 × 10 & 0.210 & 53.8 / 35.1 / 47.0 & 68.5 / 47.9 / 60.7 & TRUE\\
gait & 12 & 12 & 8 × 10 & 0.397 & 68.7 / 60.8 / 65.3 & 89.0 / 80.1 / 85.2 & FALSE\\
ECG 8-lead & 40 & 30 & 8 × 10 & 0.211 & 39.4 / 11.2 / 23.9 & 103.2 / 77.0 / 88.9 & TRUE\\
ocean & 12 & 12 & 8 × 10 & 0.362 & 59.2 / 49.8 / 60.3 & 74.3 / 58.7 / 72.1 & FALSE\\
weather & 12 & 12 & 8 × 10 & 0.407 & 29.5 / 38.6 / 41.9 & 45.4 / 60.8 / 61.5 & FALSE\\
electricity & 12 & 12 & 8 × 10 & 0.312 & 33.2 / 10.0 / 23.9 & 45.8 / 22.0 / 35.7 & FALSE\\
\bottomrule
\end{tabular}}
\end{table}

\begin{landscape}\begin{table}[H]
\centering
\caption{\label{tab:t-s3-models-3}Residual dependence and covariate diagnostics (REML residual curves): design effect DE and DE$/D$, residual correlation at one grid step and at 10\% and 25\% of the domain, share of negative correlations and the most negative one, max/min SD along $t$, principal components for 90\% of the residual variance, centered covariate rank, share of response variance removed by registration.}
\centering
\resizebox{\ifdim\width>\linewidth\linewidth\else\width\fi}{!}{
\fontsize{6}{8}\selectfont
\begin{tabular}[t]{lrrrrrrrrrrr}
\toprule
dataset & DE & DE/D & lag 1 step & lag 10\% & lag 25\% & neg. corr. share & min r & sd ratio & FPC90 & rank X & phase share\\
\midrule
ECG strain & 19.88 & 0.326 & 0.962 & 0.620 & 0.208 & 0.146 & -0.351 & 24.53 & 4 & 15 & 0.63\\
AF trial & 10.93 & 0.364 & 0.889 & 0.684 & 0.359 & 0.195 & -0.269 & 29.78 & 3 & 8 & 0.43\\
running & 33.31 & 0.546 & 0.994 & 0.871 & 0.594 & 0.119 & -0.095 & 4.63 & 2 & 7 & 0.30\\
DTI & 9.91 & 0.180 & 0.811 & 0.233 & 0.106 & 0.193 & -0.340 & 2.31 & 16 & 24 & 0.22\\
gait & 36.96 & 0.366 & 0.978 & 0.483 & 0.258 & 0.327 & -0.535 & 6.15 & 6 & 8 & 0.03\\
ECG 8-lead & 6.19 & 0.061 & 0.938 & 0.363 & 0.078 & 0.324 & -0.681 & 42.26 & 4 & 15 & 0.33\\
ocean & 39.74 & 0.393 & 0.987 & 0.738 & 0.432 & 0.217 & -0.268 & 2.86 & 5 & 6 & 0.00\\
weather & 29.12 & 0.399 & 0.503 & 0.489 & 0.393 & 0.006 & -0.101 & 2.22 & 19 & 4 & 0.06\\
electricity & 35.77 & 0.745 & 0.981 & 0.894 & 0.822 & 0.000 & 0.308 & 3.07 & 2 & 11 & 0.00\\
\bottomrule
\end{tabular}}
\end{table}
\end{landscape}

\begin{table}[H]
\centering
\caption{\label{tab:t-s3-models-4}Datasets with dependent source curves: block construction for the block flips and block-wise NCV; residual-score correlation within and between blocks.}
\centering
\resizebox{\ifdim\width>\linewidth\linewidth\else\width\fi}{!}{
\fontsize{6}{8}\selectfont
\begin{tabular}[t]{llrlrrrr}
\toprule
dataset & block rule & blocks & block size & alt. blocks & score corr. within & score corr. between & between pairs\\
\midrule
ocean & 3 consecutive monthly casts (cost check; the autocorrelation rule gives single casts) & 39 & 2–3 &  & 0.045 & -0.100 & 38\\
weather & complete-linkage clusters of station distance cut at 250 km (mismatched partition: 150 km) & 16 & 1–8 & 25 & 0.548 & 0.356 & 276\\
electricity & 2 consecutive frame curves (10 weeks) & 51 & 2 &  & 0.214 & 0.011 & 50\\
\bottomrule
\end{tabular}}
\end{table}

\subsection{Construction of the plasmode
data}\label{construction-of-the-plasmode-data}

Each dataset's model,
\(Y_i(t) = \alpha(t) + \int X_i(s)\beta(s,t)\,ds + v_i\gamma(t) + \varepsilon_i(t)\)
(Gaussian), is fitted by REML and by curve-blocked NCV. The fitted means
of three fits serve as truths: the REML fit (\emph{REML-fitted truth}),
the NCV fit (\emph{NCV-fitted truth}) and a refit with every smoothing
parameter at the geometric mean of the two fits' values (\emph{midpoint
truth}). Replicate data are these fitted means plus each subject's own
residual curve from the REML or the NCV fit, multiplied by an
independent random sign per curve. Sign flipping keeps each residual
curve's within-curve covariance and makes the curves independent across
subjects. For the ECG strain data, heartbeat difference curves
\((Y_{i1} - Y_{i2})/\sqrt 2\) of two beats per subject are a further
error source, not shaped by any fit, used on their own scale or rescaled
to the average variance of the REML residuals.

Each combination of dataset, truth and residual source is run with all
subjects and with 40 randomly chosen subjects (200 replicates each). The
three datasets with dependent source curves add block flips (one sign
per block of neighboring curves) and block-wise NCV and CL2. The mean is
evaluated at the covariates of a fixed set of 50 subjects of the
dataset, at the native grid points. There are 130 combinations in all.

\begin{table}[H]
\centering
\caption{\label{tab:t-s3-cohort}Choice of the fixed set of 50 subjects for the mean: the rule used, and the balance check of the first 50 subjects against the remaining subjects (one-sided test at 5\%, permutation p-value, standardized mean differences of the scalar covariate and the first three covariate scores). Where this check failed, a seeded random sample of 50 subjects was used; the balance of that sample was not recorded.}
\centering
\resizebox{\ifdim\width>\linewidth\linewidth\else\width\fi}{!}{
\fontsize{6}{8}\selectfont
\begin{tabular}[t]{lrllrrrrr}
\toprule
dataset & G & cohort & one-sided at 5\% & permutation p & SMD z & SMD PC1 & SMD PC2 & SMD PC3\\
\midrule
ECG strain & 78 & first 50 rows & FALSE & 0.302 & 0.313 & 0.021 & -0.180 & -0.402\\
AF trial & 118 & seeded sample of 50 & TRUE & 0.000 & 1.788 & 0.046 & -0.231 & 0.317\\
running & 90 & seeded sample of 50 & TRUE & 0.000 & -0.331 & -0.398 & 0.916 & -0.008\\
DTI & 92 & seeded sample of 50 & TRUE & 0.000 & -1.149 & -0.408 & -0.258 & -0.036\\
gait & 138 & seeded sample of 50 & TRUE & 0.000 & 1.668 & 0.118 & 0.299 & 0.377\\
ECG 8-lead & 100 & seeded sample of 50 & TRUE & 0.000 & -1.990 & -1.344 & 0.235 & 0.274\\
ocean & 116 & seeded sample of 50 & TRUE & 0.000 & -1.725 & -0.255 & -0.346 & 0.109\\
weather & 73 & first 50 rows & FALSE & 0.411 & -0.173 & 0.130 & 0.225 & 0.378\\
electricity & 102 & seeded sample of 50 & TRUE & 0.000 & -1.724 & 0.029 & -0.091 & -0.297\\
\bottomrule
\end{tabular}}
\end{table}

\section{Computing time}\label{sec-app-time}

\begin{figure}

\centering{

\pandocbounded{\includegraphics[keepaspectratio]{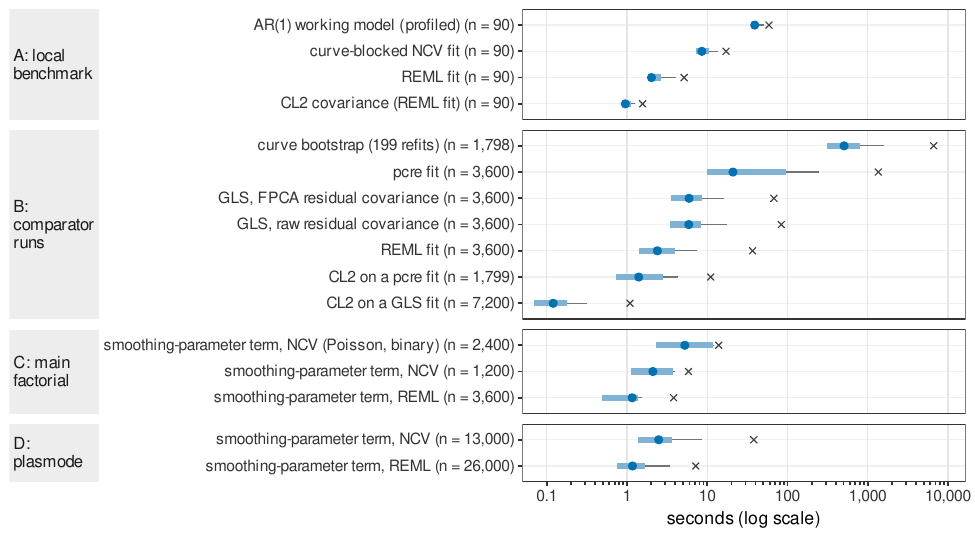}}

}

\caption{\label{fig-time}Computing time in seconds per fit or per
covariance, by method and data source (A: local benchmark, Gaussian, G =
100, D = 61; B: comparator runs, Gaussian synthetic settings and
plasmode datasets with independent source curves; C: main-factorial
settings; D: plasmode data); n: number of timed fits or covariances; log
scale. Point: median; thick bar: quartiles; thin bar: up to the 90\%
quantile; cross: maximum. CL2 and the smoothing-parameter terms are
computed on top of an existing fit. Times for pcre and GLS include the
REML pilot fit and, where used, the FPCA of its residual curves.}

\end{figure}

Times are wall-clock seconds on one core. They come from four sources:
(A) a benchmark on one local core with Gaussian data, \(G = 100\)
curves, \(D = 61\) grid points and the default bases (90 replicates over
the dependent error processes of the simulation); (B) the comparator
runs on a computing cluster, Gaussian main-factorial settings and
plasmode datasets with independent source curves, \(G = 40\) to 138; (C)
the smoothing-parameter-term runs on the main-factorial settings at the
default signal (\(G = 40\) and 100); (D) the same terms on the plasmode
data. Figure~\ref{fig-time} shows the median, the quartiles, the 90\%
quantile and the maximum per method; Supplement S7 lists them.

A curve-blocked NCV fit costs about 4.2 times a REML fit; the profiled
AR(1) model about 19 times, because it refits the model at each value of
\(\rho\). The CL2 covariance of a REML fit adds a median 0.95 s, its
Satterthwaite degrees of freedom add a median 1.8 s per replicate for
all four estimands in the main factorial, 2.1 s in the other synthetic
settings and 1.7 s on the plasmode data (cluster nodes).

Times for GLS and pcre include the REML pilot fit and the FPCA of its
residual curves (the raw GLS covariance needs no FPCA). GLS then takes a
median of 5.9 s; pcre's time grows quickly with the number of curves and
grid points (median 21 s, 90\% quantile 244 s, maximum 1356 s). CL2 on
these fits adds a median of 0.12 s (GLS) and 1.4 s (pcre). The curve
bootstrap with 199 refits takes a median of 507 s and up to 6645 s per
dataset. The smoothing-parameter uncertainty terms (curve scores,
Hessian and infinitesimal jackknife) add a median of 1.2 s to a REML fit
and 2.1 s to a Gaussian NCV fit. The one-step bias correction of the NCV
estimate needs one linear solve with \(p \times p\) matrices and was not
timed separately.

\end{document}